\documentclass{SciPost}

\hypersetup{
    colorlinks,
    linkcolor={red!50!black},
    citecolor={blue!50!black},
    urlcolor={blue!80!black}
}
\usepackage[utf8]{inputenc}
\usepackage[english]{babel}
\usepackage[T1]{fontenc}
\usepackage{amsmath,mathtools}
\usepackage{hyperref}
\usepackage{xcolor}
\usepackage{tikz}
\usepackage{lipsum}
\usepackage{bm}
\usepackage[bitstream-charter]{mathdesign}
\newcommand{\drawAMPSup}[3]{
    \begin{scope}[shift={(#1)}]
        \coordinate (origin) at (0,0);  
        
        \draw (origin) -- (1, 0);
        \draw (1, 0) -- (1, 1);
        \draw (origin) -- (0, 1);
        \draw (0, 1) -- (1, 1);
        \draw[thick, -] (-0.75, 0.5) -- (0, 0.5);
        \draw[thick, -] (1, 0.5) -- (1.75, 0.5);
        \draw[thick, -] (0.5, -0.75) -- (0.5, 0);

        \node[scale = #2] at (0.5, 0.5) {$#3$};

    \end{scope}
}

\newcommand{\drawAconjugateMPS}[3]{
    \begin{scope}[shift={(#1)}]
        \coordinate (origin) at (0,0);  
        
        \draw (origin) -- (1, 0);
        \draw (1, 0) -- (1, 1);
        \draw (origin) -- (0, 1);
        \draw (0, 1) -- (1, 1);
        \draw[thick, -]  (0, 0.5) -- (-0.75, 0.5);
        \draw[thick, -] (1.75, 0.5) -- (1, 0.5);
        \draw[thick,-] (0.5, 1) -- (0.5, 1.75);

        \node[scale = #2] at (0.5, 0.5) {$#3$};

    \end{scope}
}
\newcommand{\drawAMPS}[3]{
    \begin{scope}[shift={(#1)}]
        \coordinate (origin) at (0,0);  
        
        \draw (origin) -- (1, 0);
        \draw (1, 0) -- (1, 1);
        \draw (origin) -- (0, 1);
        \draw (0, 1) -- (1, 1);
        \draw[thick] (-0.75, 0.5) -- (0, 0.5);
        \draw[thick] (1, 0.5) -- (1.75, 0.5);
        \draw[thick] (0.5, 0) -- (0.5, -0.75);

        \node[scale = #2] at (0.5, 0.5) {$#3$};

    \end{scope}
}

\newcommand{\drawAMPSshort}[3]{
    \begin{scope}[shift={(#1)}]
        \coordinate (origin) at (0,0);  
        
        \draw (origin) -- (1, 0);
        \draw (1, 0) -- (1, 1);
        \draw (origin) -- (0, 1);
        \draw (0, 1) -- (1, 1);
        \draw[thick] (-0.75, 0.5) -- (0, 0.5);
        \draw[thick] (1, 0.5) -- (1.75, 0.5);
        \draw[thick] (0.5, 0) -- (0.5, -0.25);

        \node[scale = #2] at (0.5, 0.5) {$#3$};

    \end{scope}
}

\newcommand{\drawAconjugateMPSshort}[3]{
    \begin{scope}[shift={(#1)}]
        \coordinate (origin) at (0,0);  
        
        \draw (origin) -- (1, 0);
        \draw (1, 0) -- (1, 1);
        \draw (origin) -- (0, 1);
        \draw (0, 1) -- (1, 1);
        \draw[thick]  (0, 0.5) -- (-0.75, 0.5);
        \draw[thick] (1.75, 0.5) -- (1, 0.5);
        \draw[thick] (0.5, 1) -- (0.5, 1.25);

        \node[scale = #2] at (0.5, 0.5) {$#3$};

    \end{scope}
}

\usepackage{tikz}
\usepackage{tikz-3dplot}
\usetikzlibrary{shapes,arrows, calc, arrows.meta}
\tikzstyle{decision} = [diamond, draw, fill=blue!20, 
    text width=4.5em, text badly centered, node distance=3cm, inner sep=0pt]
\tikzstyle{block} = [rectangle, draw, fill=blue!20, 
    text width=5em, text centered, rounded corners, minimum height=4em]
\tikzstyle{line} = [draw, -latex']
\tikzstyle{cloud} = [draw, ellipse,fill=red!20, node distance=3cm,
    minimum height=2em]

\usepackage[nameinlink,noabbrev]{cleveref} 

\crefformat{equation}{#2Eq.~\textup{(\textcolor{blue}{#1})}#3}
\Crefformat{equation}{#2Equation~\textup{(\textcolor{blue}{#1})}#3}

\crefformat{figure}{#2Fig.~\textup{\textcolor{blue}{#1}}#3}
\Crefformat{figure}{#2Figure~\textup{\textcolor{blue}{#1}}#3}

\crefformat{table}{#2Table.~\textup{\textcolor{blue}{#1}}#3}
\Crefformat{table}{#2Table~\textup{\textcolor{blue}{#1}}#3}

\usepackage{caption}        
\usepackage{subcaption}     
\usepackage{comment}       

\usepackage{import}     

\DeclareSymbolFont{usualmathcal}{OMS}{cmsy}{m}{n}
\DeclareSymbolFontAlphabet{\mathcal}{usualmathcal}

\fancypagestyle{SPstyle}{
\fancyhf{}
\lhead{\colorbox{scipostblue}{\bf \color{white} ~SciPost Physics }}
\rhead{{\bf \color{scipostdeepblue} ~Submission }}

\fancyfoot[C]{\textbf{\thepage}}
}

\begin{document}

\pagestyle{SPstyle}

\begin{center}{\Large \textbf{\color{scipostdeepblue}{
Exact critical exponents of the Motzkin and Fredkin Chains\\
}}}\end{center}

\begin{center}\textbf{
Olai B. Mykland\textsuperscript{1,2,3$\star$},
 and
Zhao Zhang\textsuperscript{1$\dagger$}
}\end{center}

\begin{center}
{\bf 1} Department of Physics, University of Oslo, P.O. Box 1048 Blindern, N-0316 Oslo, Norway
\\
{\bf 2} Department of Physics, Norwegian
University of Science and Technology, NO-7491 Trondheim, Norway\\
{\bf 3}Department of Computer Science, University of Copenhagen, Copenhagen, Denmark
\\[\baselineskip]
$\star$ \href{mailto:olai.b.mykland@gmail.com }{\small olai.b.mykland@gmail.com}\,,\quad
$\dagger$ \href{mailto:zhao.zhang@fys.uio.no}{\small zhao.zhang@fys.uio.no}
\end{center}

\section*{\color{scipostdeepblue}{Abstract}}
\textbf{\boldmath{%
The Motzkin and Fredkin chains are frustration-free spin models with exactly solvable ground states whose $q$-deformations realize an exotic quantum phase transition from a disordered phase to an ordered one under domain-wall boundary conditions. Previous work has mainly focused on their entanglement scaling and spectral gaps, particularly in color-enriched variants. Here we systematically characterize the critical behavior of this transition with three main advances. First, we interpret bulk magnetic order as arising from correlations with boundary spins subject to an effective edge field, directly relating the ordered phase to the domain-wall boundary condition. Second, using the transfer matrix (TM) constructed from the exact matrix product state (MPS) representation, together with a continuum and RG analysis, we derive the algebraic decay of spin correlations and obtain the critical exponent $\eta = \frac{3}{2}$. We further generalize this TM/dual-Hamiltonian method to translationally invariant gapless states with generic power-law correlations, via a zero-dimensional Hamiltonian dual to the one-dimensional TM. Third, the TM spectrum reveals a duality between the ordered and disordered phases which, combined with scale invariance at criticality and the scaling dimension of the spin operator, yields $\nu_\pm = \frac{2}{3}$ from an RG analysis of the $q$-deformed ground states. Both exponents are confirmed numerically by MPS simulations and by direct diagonalization of the TM.
}}

\vspace{\baselineskip}

\noindent\textcolor{white!90!black}{%
\fbox{\parbox{0.975\linewidth}{%
\textcolor{white!40!black}{\begin{tabular}{lr}%
  \begin{minipage}{0.6\textwidth}%
    {\small Copyright attribution to authors. \newline
    This work is a submission to SciPost Physics. \newline
    License information to appear upon publication. \newline
    Publication information to appear upon publication.}
  \end{minipage} & \begin{minipage}{0.4\textwidth}
    {\small Received Date \newline Accepted Date \newline Published Date}%
  \end{minipage}
\end{tabular}}
}}
}


\vspace{10pt}
\noindent\rule{\textwidth}{1pt}
\tableofcontents
\noindent\rule{\textwidth}{1pt}
\vspace{10pt}


\section{Introduction}

In textbook statistical mechanics, there are two quintessential analytical tools for exact calculations. One is the transfer matrix (TM) method, which applies universally to translationally invariant one-dimensional (1D) systems. The other is the renormalization group (RG), which in higher dimensions typically relies on approximations, except on hierarchical lattices \cite{PhysRevB.58.14387}. Since quantum many-body ground states can be viewed as superpositions of classical configurations, both methods extend naturally to quantum systems, most elegantly within the framework of tensor networks (TNs).

A ground state (GS) wavefunction is a multivariate function: It takes as input the configurations of the physical degrees of freedom and outputs a complex probability amplitude. For a system of $N$ local degrees of freedom, this is a rank-$N$ tensor. To be useful for storing and manipulating information, such a tensor must be decomposed into a product of lower-rank tensors. In condensed matter physics, the systems of interest typically live on regular lattices, so the tensor decomposition should respect the discrete symmetries of the underlying lattice.

The simplest such decomposition on a 1D translationally invariant lattice is the tensor train: a contraction of rank-3 tensors arranged in a line. Rank three is special in that further decompositions only increase the number of tensors without reducing their ranks. This is precisely the structure produced by the density matrix renormalization group (DMRG) algorithm \cite{White}, except that the tensors are generally site-dependent. To fully exploit translation symmetry, one can instead start from a trial state described by a matrix product of identical rank-3 tensors, optimized variationally by algorithms such as infinite time-evolving block decimation (iTEBD) \cite{iTEBD}. The conceptual leap from DMRG as a sequential optimization scheme to matrix product states (MPS) \cite{MPS,RevModPhys.93.045003} as a general framework for gapped GSs was largely enabled by the realization that MPS give exact representations of the GSs of the AKLT chain \cite{PhysRevLett.59.799,AKLT2DCMP} and the Majumdar–Ghosh chain \cite{10.1063/1.1664978,10.1063/1.1664979}. Two-dimensional valence-bond-solid states \cite{2DAKLT} subsequently motivated the generalization of MPS to projected entangled pair states (PEPS) \cite{10.1063/1.1664979,PhysRevLett.96.220601}. The success of these constructions extended further to multicomponent simplex solid states \cite{PhysRevB.77.104404}, which inspired projected entangled simplex states \cite{PhysRevX.4.011025}.

In contrast to this MPS/PEPS development for gapped GSs, the tensor network description of gapless GSs—the multiscale entanglement renormalization ansatz (MERA)—was proposed without guidance from a specific exactly solvable GS \cite{MERAFirstPaper}. Although many GSs, especially free-fermion systems \cite{PhysRevLett.116.140403,PhysRevX.8.011003}, have been shown to be well approximated by MERA, an exactly solvable GS whose TN representation is a MERA is still lacking. Unlike MPS, which feature short-range entanglement and can be generated by finite-depth quantum circuits, critical GSs at quantum phase transitions violate the area law of entanglement entropy  logarithmically \cite{EisertAreaLaw}. This necessitates a hierarchical TN structure extending along an extra `scale' dimension beyond the physical lattice.

The simplest hierarchical TN is the tree tensor network, obtained from applying the tensor renormalization group (TRG) \cite{PhysRevLett.99.120601,PhysRevB.80.155131} to a two-dimensional TN with open boundaries. TRG, however, leaves short-range correlations intact along the RG flow and fails to accurately capture critical fixed points. To remedy this, Evenbly and Vidal introduced tensor network renormalization (TNR), which removes short-range correlations by inserting unitary disentanglers between nearest neighbors before each RG step \cite{PhysRevLett.115.180405}. The resulting MERA structure consists of rank-3 isometries and rank-4 disentanglers, but a single unit cell containing one of each can be regarded as a rank-5 tensor. The key difference between MERA and tree tensor networks, which are composed solely of rank-3 tensors, is that MERA faithfully captures entanglement at all length scales, including within each level of the coarse-grained degrees of freedom.

To construct an exactly solvable GS with genuine long-range entanglement, Bravyi et al.~\cite{PhysRevLett.109.207202} proposed a frustration-free spin-1 model, later known as the Motzkin chain. In many respects, it can be viewed as a more entangled analogue of the AKLT chain. An exact TN representation of its GS was subsequently constructed \cite{Alexander2021exactholographic}, providing the first exact TN for a critical GS. This TN shares structural features with MERA, having rank-3 and rank-4 building blocks, and exhibits the same entanglement scaling. However, the absence of isometries and unitaries makes it less amenable to the computation of correlation functions. Both TNs possess unit cells that can be regarded as rank-5 tensors, suggesting that they should be related by a sequence of singular value decompositions (SVDs) and internal contractions, i.e.~as different decompositions of a single rank-5 tensor into rank-3 and rank-4 constituents. This highlights a trade-off between informativeness and uniformity in TN representations of a given state: While any TN can be further decomposed into rank-3 tensors only, using higher-rank tensors reduces the arbitrariness in the TN definition to gauge degrees of freedom. For 1D MERA, we identify rank five as the minimal rank at which different TN representations of the same state share the same geometrical structure.

The Motzkin chain has been generalized to spin-$\frac{1}{2}$ systems with next-nearest-neighbor interactions, reminiscent of a more entangled counterpart of the Majumdar–Ghosh chain. Because the GS is a superposition of random-walk configurations, it admits a particularly simple MPS description with a bond dimension that grows with system size \cite{FredkinSpinChain}. This MPS was originally proposed as an approximate description for truncated bond dimension. Here, we restore exact translational invariance by adopting uniform tensors along the chain and use this uniform MPS to construct a TM that permits an analytical derivation of the power-law decay of correlations and the corresponding critical exponent $\eta$. To our knowledge, this is the first example in which an exact critical exponent is obtained analytically from a transfer-matrix calculation. This stands in contrast to the more common practice of numerically approximating gapless GSs with large but finite bond-dimension MPS. It is particularly striking that the same critical exponent can be extracted from the MPS representation, even though the optimized MERA structure is not as directly suited to such a calculation. By exploiting the duality between 1D TMs and zero-dimensional (0D) Hamiltonians, we show that this analytic method applies to generic gapless GSs, with the exponent $\eta$ read off directly from the spectrum of the dual 0D Hamiltonian.

The Motzkin and Fredkin models are distinguished by the fact that they do not describe a fine-tuned critical point, but rather a phase transition driven by a deformation parameter $q$, while the Hamiltonian remains frustration free \cite{ZhaoNovelPT,DeformedFredkinChain,DeformedFredkinChain_explanation,ExactRainbowTensorNetwork,ZhaoSixNineteenVertex,Zhang2024quantumlozenge,mykland2025highlyentangled2dground,zhang2025motzkinspaghetto}. Since their discovery, the associated exotic phase transition has been overshadowed in the literature by the unusual entanglement-scaling behaviors in different phases of color-enriched variants. As a consequence, it has been surprisingly overlooked as a genuine phase transition between ordered and disordered spin phases, even though the ordered phase is realized under domain-wall boundary conditions. Here we leverage both the scaling dimension of the spin operator, obtained from the TM analysis, and the self-similarity at criticality to perform an RG analysis of the $q$-deformed GSs. This allows us to compute the correlation-length exponent $\nu$. The symmetry of $\nu$ on both sides of the transition reveals a duality between the ordered and disordered phases that is not evident in the Motzkin- or Dyck-path representations.

The remainder of this article is organized as follows. In \Cref{sec:FredkinMotzkinSpinChain}, we review the Motzkin and Fredkin GSs, their parent Hamiltonians, and their common phase diagram. In \Cref{sec:transfermatrix}, we construct the TM from the MPS representation of the GS with bond dimension scaling with system size and use it to determine the spatial decay of correlations. \Cref{sec:correlation} demonstrates the power-law decay of the order parameter at criticality, reflecting its correlation with boundary spins. We then generalize the TM method in \Cref{sec:0D} to generic gapless states by introducing the dual 0D Hamiltonian, whose spectrum directly yields the critical exponent $\eta$. From the critical correlation function, we infer the scaling dimension of the spin operator, which is used in \Cref{sec:RG} to obtain the critical exponent $\nu$ via RG analysis of the $q$-deformed GSs. Both analytical exponents, $\eta$ and $\nu$, are confirmed by numerical MPS calculations. In this way, we achieve an exact characterization of the critical behavior near a phase transition in a frustration-free system by jointly exploiting translational invariance and self-similarity. Finally, \Cref{sec:conclusion} summarizes our results and outlines several directions for future work.
 
\section{Fredkin and Motzkin spin chains}
\label{sec:FredkinMotzkinSpinChain}

We begin by briefly reviewing the GSs of the deformed Fredkin and Motzkin chains, and their respective parent Hamiltonians \cite{ZhaoNovelPT,DeformedFredkinChain,DeformedFredkinChain_explanation}. The GSs of the models can be described in terms of two types of walks on the $(x, y)$ grid. The Fredkin GS is described by Dyck walks, while the Motzkin GS by Motzkin walks. A Dyck walk of $2L$ steps is defined as a path on the $(x, y)$ grid, consisting of up steps $(1, 1)$ and down steps $(1, -1)$ starting from $(0, 0)$ and ending at $(2L, 0)$, that never passes below the $x$-axis. The Motzkin walk is described by the same definition, but in addition they are diluted by flat steps, that is, also the step $(1, 0)$ is valid. Examples of Dyck walks are seen in \cref{fig:D_and_M_walks} (a) and (b) whereas all walks in \cref{fig:D_and_M_walks} are examples of Motzkin walks.

\begin{figure}[hbt!]
    \centering
    \scalebox{0.75}{
    \begin{tikzpicture}
    \begin{scope}[shift = {(2, -1.3)}, scale = 0.7]
        
        \node[scale = 1.2] at (0,0) {(a)};
        \node[scale = 1.2] at (7.5,0) {(b)};
        \node[scale = 1.2] at (14,0) {(c)};
        \node[scale = 1.2] at (22,0) {(d)};

    \end{scope}
        
    \begin{scope}[shift = {(0, 4)}, scale = 0.7]
        
        \draw[thick] (0, -4) -- (1, -3);
        \draw[thick] (1, -3) -- (2, -4);
        \draw[thick] (2, -4) -- (3, -3);
        \draw[thick] (3, -3) -- (4, -4);
        \draw[thick] (4, -4) -- (5, -3);
        \draw[thick] (5, -3) -- (6, -4);

        \draw[gray!60] (0, -4) -- (6, -4);
        
        \fill[black] (0, -4) circle (1.5pt);
        \fill[black] (1, -3) circle (1.5pt);
        \fill[black] (2, -4) circle (1.5pt);
        \fill[black] (3, -3) circle (1.5pt);
        \fill[black] (4, -4) circle (1.5pt);
        \fill[black] (5, -3) circle (1.5pt);
        \fill[black] (6, -4) circle (1.5pt);

    \end{scope}
    \begin{scope}[shift = {(0, 0)}, scale = 0.7]
        \draw [thick, arrows = {-Stealth[inset=0pt, angle=60 :4pt]}] (0.5, -0.5) -- (0.5, 0);  
        \draw [thick, arrows = {-Stealth[inset=0pt, angle=60 :4pt]}] (1.5, 0) -- (1.5, -0.5);  
        \draw [thick, arrows = {-Stealth[inset=0pt, angle=60 :4pt]}] (2.5, -0.5) -- (2.5, 0);  
        \draw [thick, arrows = {-Stealth[inset=0pt, angle=60 :4pt]}] (3.5, 0) -- (3.5, -0.5);  
        \draw [thick, arrows = {-Stealth[inset=0pt, angle=60 :4pt]}] (4.5, -0.5) -- (4.5, 0);  
        \draw [thick, arrows = {-Stealth[inset=0pt, angle=60 :4pt]}] (5.5, 0) -- (5.5, -0.5);  

        \node[scale = 0.75] at (0.5, -0.8) {$i = 1$};
        \node[scale = 0.75] at (1.5, -0.8) {$i = 2$};
        \node[scale = 0.75] at (2.5, -0.8) {$i = 3$};
        \node[scale = 0.75] at (3.5, -0.8) {$i = 4$};
        \node[scale = 0.75] at (4.5, -0.8) {$i = 5$};
        \node[scale = 0.75] at (5.5, -0.8) {$i = 6$};
        \draw[gray!60, dashed] (0.5, 0.2) -- (0.5, 5);
        \draw[gray!60, dashed] (1.5, 0.2) -- (1.5, 5);
        \draw[gray!60, dashed] (2.5, 0.2) -- (2.5, 5);
        \draw[gray!60, dashed] (3.5, 0.2) -- (3.5, 5);
        \draw[gray!60, dashed] (4.5, 0.2) -- (4.5, 5);
        \draw[gray!60, dashed] (5.5, 0.2) -- (5.5, 5);
    \end{scope}
    \begin{scope}[shift = {(5, 4)}, scale = 0.7]
        
        \draw[thick] (0, -4) -- (1, -3);
        \draw[thick] (1, -3) -- (2, -2);
        \draw[thick] (2, -2) -- (3, -1);
        \draw[thick] (3, -1) -- (4, -2);
        \draw[thick] (4, -2) -- (5, -3);
        \draw[thick] (5, -3) -- (6, -4);

        \draw[gray!60] (0, -4) -- (6, -4);
        
        \fill[black] (0, -4) circle (1.5pt);
        \fill[black] (1, -3) circle (1.5pt);
        \fill[black] (2, -2) circle (1.5pt);
        \fill[black] (3, -1) circle (1.5pt);
        \fill[black] (4, -2) circle (1.5pt);
        \fill[black] (5, -3) circle (1.5pt);
        \fill[black] (6, -4) circle (1.5pt);
    \end{scope}
    \begin{scope}[shift = {(5, 0)}, scale = 0.7]
        \draw [thick, arrows = {-Stealth[inset=0pt, angle=60 :4pt]}] (0.5, -0.5) -- (0.5, 0);  
        \draw [thick, arrows = {-Stealth[inset=0pt, angle=60 :4pt]}] (1.5, -0.5) -- (1.5, 0);  
        \draw [thick, arrows = {-Stealth[inset=0pt, angle=60 :4pt]}] (2.5, -0.5) -- (2.5, 0);  
        \draw [thick,arrows = {-Stealth[inset=0pt, angle=60 :4pt]}] (3.5, 0) -- (3.5, -0.5);  
        \draw [thick, arrows = {-Stealth[inset=0pt, angle=60 :4pt]}] (4.5, 0) -- (4.5, -0.5);  
        \draw [thick, arrows = {-Stealth[inset=0pt, angle=60 :4pt]}] (5.5, 0) -- (5.5, -0.5);  

        \node[scale = 0.75] at (0.5, -0.8) {$i = 1$};
        \node[scale = 0.75] at (1.5, -0.8) {$i = 2$};
        \node[scale = 0.75] at (2.5, -0.8) {$i = 3$};
        \node[scale = 0.75] at (3.5, -0.8) {$i = 4$};
        \node[scale = 0.75] at (4.5, -0.8) {$i = 5$};
        \node[scale = 0.75] at (5.5, -0.8) {$i = 6$};
        \draw[gray!60, dashed] (0.5, 0.2) -- (0.5, 5);
        \draw[gray!60, dashed] (1.5, 0.2) -- (1.5, 5);
        \draw[gray!60, dashed] (2.5, 0.2) -- (2.5, 5);
        \draw[gray!60, dashed] (3.5, 0.2) -- (3.5, 5);
        \draw[gray!60, dashed] (4.5, 0.2) -- (4.5, 5);
        \draw[gray!60, dashed] (5.5, 0.2) -- (5.5, 5);
    \end{scope}
    \begin{scope}[shift = {(10, 4)}, scale = 0.7]
        
        \draw[gray!60] (0, -4) -- (6, -4);
        
        \draw[thick] (0, -4) -- (1, -3);
        \draw[thick] (1, -3) -- (2, -3);
        \draw[thick] (2, -3) -- (3, -4);
        \draw[thick] (3, -4) -- (4, -4);
        \draw[thick] (4, -4) -- (5, -3);
        \draw[thick] (5, -3) -- (6, -4);

        \fill[black] (0, -4) circle (1.5pt);
        \fill[black] (1, -3) circle (1.5pt);
        \fill[black] (2, -3) circle (1.5pt);
        \fill[black] (3, -4) circle (1.5pt);
        \fill[black] (4, -4) circle (1.5pt);
        \fill[black] (5, -3) circle (1.5pt);
        \fill[black] (6, -4) circle (1.5pt);

    \end{scope}
    \begin{scope}[shift = {(10, 0)}, scale = 0.7]
        \draw [thick, arrows = {-Stealth[inset=0pt, angle=60 :4pt]}] (0.5, -0.5) -- (0.5, 0);  
        \node at (1.5, -0.25) {$0$};
        \draw [thick, arrows = {-Stealth[inset=0pt, angle=60 :4pt]}] (2.5, 0) -- (2.5, -0.5);  
        \node at (3.5, -0.25) {$0$};
        \draw [thick, arrows = {-Stealth[inset=0pt, angle=60 :4pt]}] (4.5, -0.5) -- (4.5, 0);  
        \draw [thick, arrows = {-Stealth[inset=0pt, angle=60 :4pt]}] (5.5, 0) -- (5.5, -0.5);  

        \node[scale = 0.75] at (0.5, -0.8) {$i = 1$};
        \node[scale = 0.75] at (1.5, -0.8) {$i = 2$};
        \node[scale = 0.75] at (2.5, -0.8) {$i = 3$};
        \node[scale = 0.75] at (3.5, -0.8) {$i = 4$};
        \node[scale = 0.75] at (4.5, -0.8) {$i = 5$};
        \node[scale = 0.75] at (5.5, -0.8) {$i = 6$};
        \draw[gray!60, dashed] (0.5, 0.2) -- (0.5, 5);
        \draw[gray!60, dashed] (1.5, 0.2) -- (1.5, 5);
        \draw[gray!60, dashed] (2.5, 0.2) -- (2.5, 5);
        \draw[gray!60, dashed] (3.5, 0.2) -- (3.5, 5);
        \draw[gray!60, dashed] (4.5, 0.2) -- (4.5, 5);
        \draw[gray!60, dashed] (5.5, 0.2) -- (5.5, 5);

    \end{scope}
    \begin{scope}[shift = {(15, 4)}, scale = 0.7]
            \draw[gray!60] (0, -4) -- (6, -4);
        \draw[thick] (0, -4) -- (1, -3);
        \draw[thick] (1, -3) -- (2, -2);
        \draw[thick] (2, -2) -- (3, -2);
        \draw[thick] (3, -2) -- (4, -3);
        \draw[thick] (4, -3) -- (5, -4);
        \draw[thick] (5, -4) -- (6, -4);

        \fill[black] (0, -4) circle (1.5pt);
        \fill[black] (1, -3) circle (1.5pt);
        \fill[black] (2, -2) circle (1.5pt);
        \fill[black] (3, -2) circle (1.5pt);
        \fill[black] (4, -3) circle (1.5pt);
        \fill[black] (5, -4) circle (1.5pt);
        \fill[black] (6, -4) circle (1.5pt);
    \end{scope}
    \begin{scope}[shift = {(15, 0)}, scale = 0.7]
        \draw [thick, arrows = {-Stealth[inset=0pt, angle=60 :4pt]}] (0.5, -0.5) -- (0.5, 0);  
        \draw [thick, , arrows = {-Stealth[inset=0pt, angle=60 :4pt]}] (1.5, -0.5) -- (1.5, 0);  
        \node at (2.5, -0.25) {$0$};
        \draw [thick, arrows = {-Stealth[inset=0pt, angle=60 :4pt]}] (3.5, 0) -- (3.5, -0.5);  
        \draw [thick,arrows = {-Stealth[inset=0pt, angle=60 :4pt]}] (4.5, 0) -- (4.5, -0.5);  
        \node at (5.5, -0.25) {$0$};

        \node[scale = 0.75] at (0.5, -0.8) {$i = 1$};
        \node[scale = 0.75] at (1.5, -0.8) {$i = 2$};
        \node[scale = 0.75] at (2.5, -0.8) {$i = 3$};
        \node[scale = 0.75] at (3.5, -0.8) {$i = 4$};
        \node[scale = 0.75] at (4.5, -0.8) {$i = 5$};
        \node[scale = 0.75] at (5.5, -0.8) {$i = 6$};
        \draw[gray!60, dashed] (0.5, 0.2) -- (0.5, 5);
        \draw[gray!60, dashed] (1.5, 0.2) -- (1.5, 5);
        \draw[gray!60, dashed] (2.5, 0.2) -- (2.5, 5);
        \draw[gray!60, dashed] (3.5, 0.2) -- (3.5, 5);
        \draw[gray!60, dashed] (4.5, 0.2) -- (4.5, 5);
        \draw[gray!60, dashed] (5.5, 0.2) -- (5.5, 5);
    \end{scope}

    \end{tikzpicture}
    }
    \caption{Spin configurations and corresponding Motzkin walks for $L = 3$. (a) and (b) are also Dyck walks, while (c) and (d) are not.}
    \label{fig:D_and_M_walks}
\end{figure}

Each walk is associated to a specific state in the Hilbert space $\mathcal{H}^{\otimes2L}$ of the spin chain, where $\mathcal{H}$ is the Hilbert space of each spin and $\otimes$ denotes the tensor product. For the spin-$\frac{1}{2}$ Fredkin chain (resp.~spin-$1$ Motzkin chain) $\mathcal{H}$ is 2- (resp.~3-)dimensional. Specifically, the walks are associated to eigenstates of the spin operator in the $z$ direction relating to a given spin $i$, denoted $\hat{S}^{z}_{i}$. It acts non-trivially only on the $i$th spin, that is 

\begin{equation}
\hat{S}^{z}_{i} = \hat{1}^{\otimes (i-1)}  \otimes 
{\hat{S}^{z}}
\otimes \hat{1}^{\otimes (2L-i)},
\end{equation}
where $\hat{1}$ is the identity operator. We denote the eigenstates of the operator $\hat{S}^{z}$ as $|\uparrow\rangle$ and $|\downarrow\rangle$, which has eigenvalues $\pm\frac{1}{2}$ in the Fredkin case and $\pm1$ in the Motzkin case. Additionally, we have the eigenstate $|0\rangle$ of eigenvalue $0$ in the Motzkin case. We let walks represent states of the chain by making the correspondence between a spin at position $i$ in the chain and a step taken at position $i-1/2$ in the walk in the following way
\begin{equation}
    |\uparrow\rangle_{i} \leftrightarrow (1, 1)_{i-1/2},\;\;\;\; |\downarrow\rangle_{i} \leftrightarrow (1, -1)_{i-1/2},\;\;\;\;|0\rangle_{i} \leftrightarrow (1, 0)_{i-1/2}. 
    \label{eq: walk_state_correspondence}
\end{equation}
This correspondence is seen in \cref{fig:D_and_M_walks}, where the subscript $i-1/2$ indicate the midpoint between spins, marked by black dots. Clearly, states described by walks using the correspondence \cref{eq: walk_state_correspondence} is a mutual eigenstate of all operators $\hat{S}^{z}_{i}$, meaning that 

\begin{equation}
    \hat{S}^{z}_{i}|w \rangle = c_i|w\rangle, \;\forall i,
\end{equation}
where $c = \pm \frac{1}{2}$ for the Fredkin chains and $c = 0, \pm1$ for the Motzkin chain. We call such states $|w\rangle$ spin configurations. \\

Using the correspondence \cref{eq: walk_state_correspondence}, the unique GSs of the deformed Fredkin and Motzkin chain $|\text{GS}(q)\rangle$, can be expressed as

\begin{equation}
    |\text{GS}(q)\rangle = \frac{1}{\sqrt{C}}\sum_{w\in D_{2L}}q^{A(w)}|w\rangle,
    \label{eq: GSDeformed}
\end{equation}
where $C$ is a normalization constant and $D_{2L}$ is the set of all Dyck or Motzkin walks on $2L$ steps, depending on which model we consider. Note also that $A(w)$ is the total area beneath the walk $w$ and that $q$ is the deformation parameter. Clearly, at $q = 1$ all walks are weighted equally, but at $q>1$ (resp.~$q<1$) high area walks (resp.~low area walks) are weighted more. \\

To understand the unusual entanglement properties of the state in \cref{eq: GSDeformed}, one can perform a Schmidt decomposition at the midpoint of the spin chain. Following \cite{DeformedFredkinChain}, this gives

\begin{equation}
    \begin{split}
    |\text{GS}(q)\rangle =& \sum^{d^{L}}_{m = 1}\alpha_{m}|\psi^{1, ..., L}_{m}\rangle\otimes|\psi^{L+1, ..., 2L}_{m}\rangle,\\
    =& \sum^{L}_{m=0}\sqrt{p_{L, m}(q)}|C^{1, ..., L}_{0, m}\rangle\otimes|C^{L+1, ..., 2L}_{m, 0}\rangle,
    \end{split}
    \label{eq: SchmidtDecompositionDeformedFredkin}
\end{equation}
where $d$ is the dimension of the Hilbert space $\mathcal{H}$ of each spin, with $d = 2$ (resp.~$d = 3$) for the Fredkin (resp.~Motzkin) chain. Note that the states $|C^{i,..., j}_{a, b}\rangle$ are area weighted superpositions of spin configurations with either $a$ excess down spins or $b$ excess up spins on the spins $i$ to $j$. In \cite{PhysRevLett.109.207202, FredkinSpinChain, ZhaoNovelPT, DeformedFredkinChain}, analysis of the Schmidt coefficients $\sqrt{p_{L, m}(q)}$ were made, which revealed that the GSs undergo a quantum criticality at the $q = 1$ point. The GSs satisfy the area law of EE both for the ordered phase $q>1$ and the disordered $q<1$ phase, but have logarithmically scaling entanglement entropy at the critical point $q = 1$. Our TM approach in \Cref{sec:transfermatrix} and \Cref{sec:correlation} combined with the RG analysis in \Cref{sec:RG} showed that both the ordered and disordered phases have exponentially decaying correlations. On the other hand, the gapless critical point exhibits power-law decay of correlations. These results are summarized in the phase diagram in \cref{fig: FredkinPhaseDiagram}. Note that the exponential decay of correlations in the ordered phase does not imply a spectral gap of the Hamiltonian. In fact, Ref.~\cite{DeformedFredkinChain_explanation} has established an exponentially small upper bound on the spectral gap. This is due to the boundary magnetic field in the Hamiltonian lifting the otherwise exponentially degenerate GSs with different magnetizations to a unique GS with $S^z_\mathrm{tot}=0$. This boundary effect turns out not to be important for the disordered phase, so the $q<1$ phase is indeed gapped \cite{andrei2022spinonemotzkinchaingapped}.
\begin{figure}[hbt!]
    \centering
    \begin{tikzpicture}
        \draw[thick, ->] (0, -1) -- (10, -1);
        \node at (10.1, -1.2) {$q$};
        \draw[dashed] (5, -1.5) -- (5, -0.9);
        \node at (5, -1.7) {$q = 1$};
        \node at (2.5, -1.3) {disordered, $S_{2L} \propto \mathcal{O}(1)$};
        \node at (5, -0.7) {gapless, $S_{2L} \propto \mathcal{O}(\log L)$}; 
        \node at (7.5, -1.3) {ordered, $S_{2L} \propto \mathcal{O}(1)$};
    \end{tikzpicture}
    \caption{Phase diagram of the deformed Fredkin and Motzkin spin chain, showing how the entanglement entropy $S_{2L}$ for the half chain partition of spin chain of length $2L$, undergoes a phase transition at the critical point $q=1$. The entanglement entropy is bounded for both $q>1$ and $q<1$, but shows a logarithmic violation of the area law at $q=1$.}
    \label{fig: FredkinPhaseDiagram}
\end{figure}

The Fredkin and Motzkin Hamiltonians were introduced in Ref.~\cite{FredkinSpinChain} and Ref.~\cite{PhysRevLett.109.207202}, respectively, and later extended to deformed versions in Ref.~\cite{DeformedFredkinChain} and Ref.~\cite{ZhaoNovelPT}. Both can be written as a sum of projectors onto states expressed in terms of the eigenstates of the single-site $\hat{S}^{z}$ operator. The Hamiltonians is written as 

\begin{equation}
    H^{\text{F/M}}_{\text{tot}}(q) = |\downarrow\rangle_{1}\langle\downarrow| + |\uparrow\rangle_{2L}\langle\uparrow|  +  \sum_{i = 2}\Pi^{\text{F/M}}_{i}(q),
    \label{eq: DeformedFredkinMotzkinH}
\end{equation}
where $q$ is the deformation parameter and the sum over $\Pi^{\text{F/M}}_{i}(q)$ terminates at $i = 2L-1$ (resp.~$i=2L$) for the Fredkin (resp.~Motzkin) Hamiltonian. The projectors $\Pi^{\text{F/M}}_{i}(q)$ are defined as

\begin{equation}
        \Pi^{\text{F}}_{i}(q) = |F^{1}_{i}\rangle \langle F^{1}_{i}| + |F^{2}_{i}\rangle \langle F^{2}_{i}|,\;\;\;\Pi^{\text{M}}_{i}(q) = |M^{1}_{i}\rangle \langle M^{1}_{i}| + |M^{2}_{i}\rangle \langle M^{2}_{i}|+ |M^{3}_{i}\rangle \langle M^{3}_{i}|.
    \label{eq: ProjectionOperator}
\end{equation}
The $q$-dependent state vectors in the Fredkin projector $\Pi^{\text{F}}_{i}(q)$ are defined as 

\begin{equation}
    \begin{split}
        |F^{1}_{i}\rangle =& \frac{1}{\sqrt{q^{-2}+q^{2}}}\left(q^{-1}|\uparrow_{i-1}\uparrow_{i}\downarrow_{i+1}\rangle - q|\uparrow_{i-1}\downarrow_{i}\uparrow_{i+1}\rangle\right),\\
        |F^{2}_{i}\rangle =& \frac{1}{\sqrt{q^{-2}+q^{2}}}\left(q^{-1}|\uparrow_{i-1}\downarrow_{i}\downarrow_{i+1}\rangle - q|\downarrow_{i-1}\uparrow_{i}\downarrow_{i+1}\rangle\right),
    \end{split}
    \label{eq: FredkinStates}
\end{equation} whereas the $q$-dependent state vectors in the Motzkin projector are defined as 

\begin{equation}
    \begin{split}
        |M^{1}_{i}\rangle =& \frac{1}{\sqrt{1+q^2}}\left(|\uparrow_{i-1}0_{i}\rangle - q|0_{i-1}\uparrow_{i}\rangle\right),\\
        |M^{2}_{i}\rangle =& \frac{1}{\sqrt{1+q^2}}\left(|0_{i-1}\downarrow_{i}\rangle - q|\downarrow_{i-1}0_{i}\rangle\right),\\
        |M^{3}_{i}\rangle =& \frac{1}{\sqrt{1+q^2}}\left(|\uparrow_{i-1}\downarrow{i}\rangle- q|0_{i-1}0_{i}\rangle \right).
    \end{split}
    \label{eq: MotzkinStates}
\end{equation}
The states in the superpositions in $|F^{j}_{i}\rangle$ and $|M^{j}_{i}\rangle$ are related by the so-called Fredkin moves $F^{j}_{i}$ and Motzkin moves $M^{j}_{i}$, seen in \cref{fig:FredkinMotzkinMoves}. Note that the Fredkin projectors involve three spins, so the model includes next-nearest-neighbor interactions. Motzkin projectors involve only two spins, so the interactions are limited to nearest neighbors. Also note the $q$-factors in \cref{eq: FredkinStates} and \cref{eq: MotzkinStates} are different. This can be understood in terms of the walks associated to the spin configurations, as seen in \cref{fig:FredkinMotzkinMoves}. Clearly, the area under the walk segments differs by 2 units for the Fredkin case but only 1 unit for the Motzkin case. In order to ensure that the weighting of the spin configurations in \cref{eq: GSDeformed} gives the correct GS, the low-area spin configurations are weighted with a factor $q^{2}$ (resp.~$q$) more than the high-area spin configuration in the Fredkin case (resp.~Motzkin case), in \cref{eq: FredkinStates} and \cref{eq: MotzkinStates}. The Hamiltonians $H^{\text{F/M}}_{\text{tot}}$ are frustration free, meaning that the GSs are the lowest energy eigenstate of each individual term in \cref{eq: DeformedFredkinMotzkinH}, with GS energy 0. That is, the GSs are annihilated by each individual projector.

\begin{figure}[hbt!]
    \centering
    \scalebox{0.7}{
    \begin{tikzpicture}
        \def\horizontalSpacing{3.8} 
        \def\verticalSpacing{1}   
        \begin{scope}[shift = {(0, 0)}]

            \draw[thin, gray!50] (-0.25, 2) -- (19, 2);
            \node at (-0.8, 2) {$h$};
            
            \draw[thin, gray!50] (-0.25, 1) -- (19, 1);
            \node at (-0.8, 1) {$h-1$};

            \draw[thin, gray!50] (-0.25, -1.5) -- (19, -1.5);
            \node at (-0.8, -1.5) {$h$};
            
            \draw[thin, gray!50] (-0.25, -2.5) -- (19, -2.5);
            \node at (-0.8, -2.5) {$h-1$};

        \end{scope}
    
        \begin{scope}[shift={(0, \verticalSpacing/2)}] 
            \draw[thick] (0, 0.5) -- (1, 1.5);
            \draw[thick] (1, 1.5) -- (2, 2.5);
            \draw[thick] (2, 2.5) -- (3, 1.5);
            \draw[gray!60, dashed] (0.5, 0) -- (0.5, 2.5);
            \draw[gray!60, dashed] (1.5, 0) -- (1.5, 2.5);
            \draw[gray!60, dashed] (2.5, 0) -- (2.5, 2.5);

            \fill[black] (1, 1.5) circle (1.5pt);
            \fill[black] (2, 2.5) circle (1.5pt);
            \fill[black] (3, 1.5) circle (1.5pt);
            \fill[black] (0, 0.5) circle (1.5pt);

            \node at (0.5,-0.5) {$i-1$};
            \node at (1.5,-0.5) {$i$};
            \node at (2.5,-0.5) {$i+1$};
            \draw [thick, arrows = {-Stealth[inset=0pt, angle=60 :4pt]}] (0.5, -0.3) -- (0.5, 0.2);  
            \draw [thick, arrows = {-Stealth[inset=0pt, angle=60 :4pt]}] (1.5, -0.3) -- (1.5, 0.2);  
            \draw [thick, arrows = {-Stealth[inset=0pt, angle=60 :4pt]}] (2.5, 0.2) -- (2.5, -0.3);  
            
        \end{scope}

        \node[scale = 1.8] at (4, 2) {$\leftrightarrow$};
        \node at (4, 2.5) {$F^{1}_{i}$};
    
        \begin{scope}[shift={(1.25*\horizontalSpacing, \verticalSpacing/2)}] 
            \draw[thick] (0, 0.5) -- (1, 1.5);
            \draw[thick] (1, 1.5) -- (2, 0.5);
            \draw[thick] (2, 0.5) -- (3, 1.5);
            \draw[gray!60, dashed] (0.5, 0) -- (0.5, 2.5);
            \draw[gray!60, dashed] (1.5, 0) -- (1.5, 2.5);
            \draw[gray!60, dashed] (2.5, 0) -- (2.5, 2.5);
    
            \fill[black] (0, 0.5) circle (1.5pt);
            \fill[black] (1, 1.5) circle (1.5pt);
            \fill[black] (2, 0.5) circle (1.5pt);
            \fill[black] (3, 1.5) circle (1.5pt);

            \node at (0.5,-0.5) {$i-1$};
            \node at (1.5,-0.5) {$i$};
            \node at (2.5,-0.5) {$i+1$};

            \draw [thick, arrows = {-Stealth[inset=0pt, angle=60 :4pt]}] (0.5, -0.3) -- (0.5, 0.2);  
            \draw [thick, arrows = {-Stealth[inset=0pt, angle=60 :4pt]}] (1.5, 0.2) -- (1.5, -0.3);  
            \draw [thick, arrows = {-Stealth[inset=0pt, angle=60 :4pt]}] (2.5, -0.3) -- (2.5, 0.2);  
)
        \end{scope}
    
        \begin{scope}[shift={(2.75*\horizontalSpacing, 3.5*\verticalSpacing)}] 
            \draw[thick] (0, -1.5) -- (1, -0.5);
            \draw[thick] (1, -0.5) -- (2, -1.5);
            \draw[thick] (2, -1.5) -- (3, -2.5);
            \draw[gray!60, dashed] (0.5, -2.5) -- (0.5, 0);
            \draw[gray!60, dashed] (1.5, 0) -- (1.5, -2.5);
            \draw[gray!60, dashed] (2.5, -2.5) -- (2.5, 0);
    
            \fill[black] (0, -1.5) circle (1.5pt);
            \fill[black] (1, -0.5) circle (1.5pt);
            \fill[black] (2, -1.5) circle (1.5pt);
            \fill[black] (3, -2.5) circle (1.5pt);

            \node at (0.5,-3.5) {$i-1$};
            \node at (1.5,-3.5) {$i$};
            \node at (2.5,-3.5) {$i+1$};

            \draw [thick, arrows = {-Stealth[inset=0pt, angle=60 :4pt]}] (0.5, -3.3) -- (0.5, -2.8);  
            \draw [thick, arrows = {-Stealth[inset=0pt, angle=60 :4pt]}] (1.5, -2.8) -- (1.5, -3.3);  
            \draw [thick, arrows = {-Stealth[inset=0pt, angle=60 :4pt]}] (2.5, -2.8) -- (2.5, -3.3);  
        \end{scope}

        \node[scale = 1.8] at (14.25, 2) {$\leftrightarrow$};
        \node at (14.25, 2.5) {$F^{2}_{i}$};
    
        \begin{scope}[shift={(4.2*\horizontalSpacing, 3.5*\verticalSpacing)}] 
            \draw[thick] (0, -1.5) -- (1, -2.5);
            \draw[thick] (1, -2.5) -- (2, -1.5);
            \draw[thick] (2, -1.5) -- (3, -2.5);
            \draw[gray!60, dashed] (0.5, -2.5) -- (0.5, 0);
            \draw[gray!60, dashed] (1.5, 0) -- (1.5, -2.5);
            \draw[gray!60, dashed] (2.5, -2.5) -- (2.5, 0);
    
            \fill[black] (0, -1.5) circle (1.5pt);
            \fill[black] (1, -2.5) circle (1.5pt);
            \fill[black] (2, -1.5) circle (1.5pt);
            \fill[black] (3, -2.5) circle (1.5pt);

            \node at (0.5,-3.5) {$i-1$};
            \node at (1.5,-3.5) {$i$};
            \node at (2.5,-3.5) {$i+1$};

            \draw [thick, arrows = {-Stealth[inset=0pt, angle=60 :4pt]}] (0.5, -2.8) -- (0.5, -3.3);  
            \draw [thick, arrows = {-Stealth[inset=0pt, angle=60 :4pt]}] (1.5, -3.3) -- (1.5, -2.8);  
            \draw [thick, arrows = {-Stealth[inset=0pt, angle=60 :4pt]}] (2.5, -2.8) -- (2.5, -3.3);  
        \end{scope}

            \begin{scope}[shift={(0, -3*\verticalSpacing)}] 
            \draw[thick] (0, 0.5) -- (1, 1.5);
            \draw[thick] (1, 1.5) -- (2, 1.5);
            \draw[gray!60, dashed] (0.5, 0) -- (0.5, 1.6);
            \draw[gray!60, dashed] (1.5, 0) -- (1.5, 1.6);

            \fill[black] (0, 0.5) circle (1.5pt);
            \fill[black] (1, 1.5) circle (1.5pt);
            \fill[black] (2, 1.5) circle (1.5pt);

            \node at (0.5,-0.5) {$i-1$};
            \node at (1.5,-0.5) {$i$};

            \draw [thick, arrows = {-Stealth[inset=0pt, angle=60 :4pt]}] (0.5, -0.3) -- (0.5, 0.2);  
            \node at (1.5, -0.05) {$0$};  

        \end{scope}
        
        \node[scale = 1.8] at (2.75, -1.75) {$\leftrightarrow$};
        \node at (2.75, -1.25) {$M^{1}_{i}$};
        
        \begin{scope}[shift={(0.85*\horizontalSpacing, -3*\verticalSpacing)}] 
            \draw[thick] (0, 0.5) -- (1, 0.5);
            \draw[thick] (1, 0.5) -- (2, 1.5);
            \draw[gray!60, dashed] (0.5, 0) -- (0.5, 1.6);
            \draw[gray!60, dashed] (1.5, 0) -- (1.5, 1.6);

            \fill[black] (0, 0.5) circle (1.5pt);
            \fill[black] (1, 0.5) circle (1.5pt);
            \fill[black] (2, 1.5) circle (1.5pt);

            \node at (0.5,-0.5) {$i-1$};
            \node at (1.5,-0.5) {$i$};

            \draw [thick, arrows = {-Stealth[inset=0pt, angle=60 :4pt]}] (1.5, -0.3) -- (1.5, 0.2);  
            \node at (0.5, -0.05) {$0$};  

        \end{scope}
        
        \begin{scope}[shift={(1.8*\horizontalSpacing, -3*\verticalSpacing)}] 
            \draw[thick] (0, 1.5) -- (1, 1.5);
            \draw[thick] (1,1.5) -- (2, 0.5);
            \draw[gray!60, dashed] (0.5, 0) -- (0.5, 1.6);
            \draw[gray!60, dashed] (1.5, 0) -- (1.5, 1.6);

            \fill[black] (0, 1.5) circle (1.5pt);
            \fill[black] (1, 1.5) circle (1.5pt);
            \fill[black] (2, 0.5) circle (1.5pt);

            \node at (0.5,-0.5) {$i-1$};
            \node at (1.5,-0.5) {$i$};

            \draw [thick, arrows = {-Stealth[inset=0pt, angle=60 :4pt]}] (1.5, 0.2) -- (1.5, -0.3);  
            \node at (0.5, -0.05) {$0$};  

        \end{scope}
        
        \node[scale = 1.8] at (9.5, -1.75) {$\leftrightarrow$};
        \node at (9.5, -1.25) {$M^{2}_{i}$};

        \begin{scope}[shift={(2.75*\horizontalSpacing, -3*\verticalSpacing)}] 
            \draw[thick] (0, 1.5) -- (1, 0.5);
            \draw[thick] (1,0.5) -- (2, 0.5);
            \draw[gray!60, dashed] (0.5, 0) -- (0.5, 1.6);
            \draw[gray!60, dashed] (1.5, 0) -- (1.5, 1.6);

            \fill[black] (0, 1.5) circle (1.5pt);
            \fill[black] (1, 0.5) circle (1.5pt);
            \fill[black] (2, 0.5) circle (1.5pt);

            \node at (0.5,-0.5) {$i-1$};
            \node at (1.5,-0.5) {$i$};

            \draw [thick, arrows = {-Stealth[inset=0pt, angle=60 :4pt]}] (0.5, 0.2) -- (0.5, -0.3);  
            \node at (1.5, -0.05) {$0$};  

        \end{scope}
        \begin{scope}[shift={(3.6*\horizontalSpacing, -3*\verticalSpacing)}] 
            \draw[thick] (0, 0.5) -- (1, 1.5);
            \draw[thick] (1,1.5) -- (2, 0.5);
            \draw[gray!60, dashed] (0.5, 0) -- (0.5, 1.6);
            \draw[gray!60, dashed] (1.5, 0) -- (1.5, 1.6);

            \fill[black] (0, 0.5) circle (1.5pt);
            \fill[black] (1, 1.5) circle (1.5pt);
            \fill[black] (2, 0.5) circle (1.5pt);

            \node at (0.5,-0.5) {$i-1$};
            \node at (1.5,-0.5) {$i$};

            \draw [thick, arrows = {-Stealth[inset=0pt, angle=60 :4pt]}] (0.5, -0.3) -- (0.5, 0.2);  
            \draw [thick, arrows = {-Stealth[inset=0pt, angle=60 :4pt]}] (1.5, 0.2) -- (1.5, -0.3);

        \end{scope}
        
        \node[scale = 1.8] at (16.2, -1.75) {$\leftrightarrow$};
        \node at (16.2, -1.25) {$M^{3}_{i}$};

        \begin{scope}[shift={(4.45*\horizontalSpacing, -3*\verticalSpacing)}] 
            \draw[thick] (0, 0.5) -- (1, 0.5);
            \draw[thick] (0.5,0.5) -- (2, 0.5);
            \draw[gray!60, dashed] (0.5, 0) -- (0.5, 1.6);
            \draw[gray!60, dashed] (1.5, 0) -- (1.5, 1.6);

            \fill[black] (0, 0.5) circle (1.5pt);
            \fill[black] (1, 0.5) circle (1.5pt);
            \fill[black] (2, 0.5) circle (1.5pt);

            \node at (0.5,-0.5) {$i-1$};
            \node at (1.5,-0.5) {$i$};

            \node at (1.5, -0.05) {$0$};  
            \node at (0.5, -0.05) {$0$};

        \end{scope}

    \end{tikzpicture}
    }

    \caption{Upper panel: the two Fredkin moves $F^{1}_{i}$ and $F^{1}_{i}$ related to the states in \cref{eq: FredkinStates}. Moving from left to right reduces the area under the walk segment by 2 units. Lower panel: the three Motzkin moves $M^{1}_{i}$, $M^{2}_{i}$ and $M^{2}_{i}$ related to the states in \cref{eq: MotzkinStates}. Moving from left to right reduces the area under the walk segment by 1 unit. The solid gray lines indicate the height.}
    \label{fig:FredkinMotzkinMoves}
\end{figure}

The fact that $H^{\text{F/M}}_{\text{tot}}$ give the GS in \cref{eq: GSDeformed} can be understood in terms of the Fredkin and Motzkin moves. First note that the boundary terms in \cref{eq: DeformedFredkinMotzkinH} penalizes any spin configuration with a down spin at $i = 1$ and/or an up spin at $i = 2L$, with an energy contribution. Thus, in order for a state to be annihilated by the boundary projectors in \cref{eq: DeformedFredkinMotzkinH}, none of the spin configuration present in the state can start with a down spin and/or end with an up spin. Next, to construct a state that is annihilated by all bulk terms $\Pi^{\text{F/M}}_{i}(q)$, one can start from a given spin configuration and then include spin configurations related by a Fredkin or Motzkin move at $i$ in the state. By doing this for all $i$ and for all included spin configurations, we can assure that each $\Pi^{\text{F/M}}_{i}(q)$ is annihilated by weighting the spin configurations correctly. As previously mentioned, the correct weighting is determined by looking at \cref{fig:FredkinMotzkinMoves}, where Fredkin (resp.~Motzkin) moves relate spin configurations that differ in 2 area units (resp.~1 area unit) under the line segments traced out by the walk. In \cref{eq: FredkinStates} and \cref{eq: MotzkinStates}, we have weighted the low-area configurations with a factor $q^{2}$ (resp.~$q$) more than the high-area configuration in the Fredkin case (resp.~Motzkin case). That is, the low-area configuration is weighted a factor $q^{\Delta A}$ more, where $\Delta A$ is the difference in area under the walks. This implies that the correct weighting of the configurations in the GS for the projector $\Pi^{\text{F/M}}_{i}(q)$ to be annihilated is to weight the high-area configuration a factor of $q^{\Delta A}$ more than the low area configuration. This is achieved by weighting each Dyck walk $w$ by a factor $q^{A(w)}$, where $A(w)$ is the area beneath the walk $w$. This is exactly what is done in the GS in \cref{eq: GSDeformed}.

Finally, note that by applying a Fredkin move $F^{i}_{j}$ (resp.~Motzkin move $M^{j}_{i}$) to a spin configuration described by a Dyck walk (resp.~Motzkin walk), one always gets another Dyck walk (resp.~Motzkin walk). Starting from a spin configuration and including all spin configurations related by a series of Fredkin moves, will always lead to a state annihilated by all $\Pi^{\text{F/M}}_{i}(q)$, but only if the initial spin configuration is described by a Dyck walk (resp.~Motzkin walk) will this procedure lead to a state that is annihilated by the boundary terms in \cref{eq: DeformedFredkinMotzkinH}. For example, if we start from a spin configuration described by a walk that reaches negative height at some point, successive Fredkin or Motzkin moves will show that we are forced to include a spin configuration where the spin at $i = 1$ is a down spin in order to make sure that the projector $\Pi^{\text{F/M}}_{2}(q)$ is annihilated. Similarly, if we start from a state described by a walk that ends at a positive height, successive Fredkin moves (resp.~Motzkin moves) shows that we are forced to include a state with an up spin at $i = 2L$ in order to make sure $\Pi^{\text{F}}_{2L-1 }(q)$ (resp.~$\Pi^{\text{M}}_{2L}(q)$) is annihilated. In total, this means that the GSs of the models are given by the are weighted superposition of Dyck or Motzkin walks in \cref{eq: GSDeformed}.

\section{MPS representations and transfer matrices}\label{sec:transfermatrix}

The Motzkin and Fredkin GSs reviewed in the previous section are special in that they can be described by two completely different TN representations. One of them has a hierarchical structure resembling the MERA as one would expect from a TN representation for critical gapless GSs, but without unitarity and isometry of the consisting tensors \cite{Alexander2021exactholographic}. In \Cref{sec:penta}, we unify the hierarchical TN and MERA to a hierarchical TN composed of one rank-five tensors. Due to the lack of unitarity, the generalized MERA do not generically facilitate efficient evaluation of the expectation values and correlation functions of local operators as the original MERA does \cite{PhysRevB.79.144108}. However, here we present an alternative analytical method to determine the power-law decay behavior of correlation functions from the second TN representation of the same GSs, which are translationally invariant \cite{ExactRainbowTensorNetwork}.

The advantage of the translationally invariant 2D TNs, where the tensors are arranged on a rectangular lattice, is that they are also compatible with the $q$-deformed and even colored counterparts of the Motzkin and Fredkin chains. From these representations, one can obtain the MPS representations by contracting a tower of tensors along the vertical direction. Naively, this would result in an MPS with exponentially large bond dimension, but the sparse tensor can be compressed to one with bond dimension scaling linearly with the system size, using the height representation of the GSs, see also Ref. \cite{mykland2025highlyentangled2dground}. Recently, the reverse process, namely decomposing a MPS tensor into the contraction of a tower of tensor in a 2D bulk, has been used to construct a holographic TN representation for the asymmetric simple exclusion process and double-scaled Sachdev-Ye-Kitaev model \cite{10.21468/SciPostPhys.19.4.083}.  

In theory, MPSs with finite bond dimension always give exponential decay of correlations, which describes 1D gapped systems. Nevertheless, it is common practice for numerical experts to use MPSs with large bond dimensions to simulate also 1D gapless states. Exceptionally, in this case, we can use the exact TMs constructed from the infinite dimensional MPS representations to analytically obtain a power law decay of the correlation. Furthermore, the MPS also describes the two gapped phases as the parameter $q$ is deformed away from its critical value 1, allowing us to study numerically the critical behavior near the phase transition reviewed in \Cref{sec:FredkinMotzkinSpinChain}. In this section, we prepare for the explicit calculation of the critical exponents in the next two sections by explicitly constructing the TMs from MPS representations.

The expectation value of spins and their correlations have been obtained before using asymptotics of combinatorial results \cite{10.1063/1.4977829, Menon:2024vic} and Gaussian free field description in the scaling limit \cite{mykland2025highlyentangled2dground}. Here we present a much more transparent derivation using the TM method, from which the critical exponents $\eta,\nu$ defined by
\begin{equation}
     \langle O_1O_r\rangle\propto \begin{cases}
         r^{2-d-\eta}, \ \mathrm{for}\ q=1,\\ e^{-r/\xi}, \quad \mathrm{for}\ q< 1,\ \mathrm{with} \ \xi\propto (-\log q)^{-\nu}
     \end{cases} \label{eq:criticalexponent}
\end{equation}
can be extracted. Due to the opposite boundary conditions on different ends of the chain, the two-point correlation in the ordered phase $q>1$ is not a constant. This will be discussed in detail in \Cref{sec:RG}. The Hamiltonians of the Motzkin and Fredkin chains (reviewed in \Cref{sec:FredkinMotzkinSpinChain}) explicitly break the symmetry between up and down spins at the boundary of a finite sized system. Hence the correlation between the spin at distance $r$ with the constant valued boundary spin in the GS can be probed from the skin-depth of the boundary effect, instead of the two-point correlation function as defined in \cref{eq:criticalexponent}. Thus, we can find the critical exponents by computing the one-point function $\langle S^{z}_{r}\rangle$.

The exact MPS representation of the Fredkin GS was given in Ref.~\cite{FredkinSpinChain} at the critical value $q = 1$ of the deformation parameter $q$. In the MPS, the bond dimension of internal legs is $L+1$ for a chain of length $2L$, which follows from the maximum height of a Dyck walk of $2L$ steps. The $L$-dependent bond dimension enable the MPS to capture the unusual scaling of the EE in the state, which scales as $\log L$. We now construct the MPS representation of the Motzkin GS, which is done by simply considering the Motzkin walks instead of the Dyck walks. We also generalize both MPSs to represent the GSs at all values of $q$. This is achieved by weighing tensor elements by factors of $q$ that depend on the height $h$ (or equivalently the area $A(w)$) of the walk $w$ in the GS superposition \cref{eq: GSDeformed}. The GS is then represented as
\begin{equation}
    \begin{tikzpicture} 
        \fill[white] (0,1) circle (1pt); 
        \node at (0, 0) {$|\mathrm{GS}\rangle=\displaystyle\sum_{\{s_j \}}A_{0h_2}^{(s_1)}A_{h_2h_3}^{(s_2)}\cdots A_{h_{2L}0}^{(s_{2L})}|\{s_j\}\rangle=$};
            \begin{scope}[scale = 0.6, shift = {(4.25, 0.13)}]
                \fill[black] (1.1, 0.2) circle (5pt);
                \drawAMPS{2, -0.3, 0}{0.7}{A^{(s_{1})}} 
                \drawAMPS{4, -0.3, 0}{0.7}{A^{(s_{2})}} 
                \node at (6.25, 0.2) {$\dots$};
            \end{scope}
            \begin{scope}[scale = 0.6, shift = {(3.5, 0)}]
                \drawAMPS{9.2, -0.3, 0}{0.6}{A^{(s_{2L})}} 
                \fill[black] (11.1, 0.2) circle (5pt);
            \end{scope}.
        \end{tikzpicture}
\label{eq: MPS}
\end{equation}
In the first expression, Einstein's summation convention applies to the internal indices $h_j$. The physical spin indices are denoted $s_{j}$ and its values corresponds to the eigenstates of $S^{z}_{j}$ for site $j$. Naturally, for the Fredkin MPS it takes on two values $s_{j} = [-1, 1]$, while for the Motzkin MPS it takes on three values $s_{j} = [-1, 0, 1]$. In the second expression, we have written the MPS diagrammatically, where each open leg corresponds to a spin $s_{j}$ and the black dots denote boundary tensors $\delta_{h_{i}0}$, fixing the indices $h_{1}$ and $h_{2L+1}$ to $0$. The matrices $A^{(s_{j})}$ are defined as
\begin{equation}
        A_{h_jh_{j+1}}^{(s_{j} )}=q^{h_j+\frac{s_{j}}{2}}\delta_{h_j,h_{j+1}+ s_{j}}.
\label{eq: MPS_Motzkin_tensor}
\end{equation}
We have $ 0\le h_j\le L$, meaning that the bond dimension of the MPS is $L+1$. Note that this definition of the MPS-tensor, implies that $|\text{GS}\rangle$ in \cref{eq: MPS} is unnormalized. The MPS is implemented in ITensor \cite{Fishman_2022} for both the Fredkin and Motzkin GSs and used to compute expectation values. The results will be presented in the next two sections.\\

The TM $\mathcal{T}$ is constructed by contracting the physical external legs of matrices $A^{(s_{j})}$ and their conjugate $\bar{A}^{(s_{j})}=A^{(s_{j})}$
\begin{equation}
\begin{split}
    &\mathcal{T}_{h_j,h_j';h_{j+1},h_{j+1}'}=\sum_{s_{j}=\pm 1 (,0) } A^{(s_{j})}_{h_j,h_{j+1}} \bar{A}^{(s_{j})}_{h_j',h_{j+1}'}\\
    &=q^{h_{j} + h^{\prime}_{j}}\left[q\delta_{h_j,h_{j+1}+1}\delta_{h'_j,h'_{j+1}+1} +q^{-1}\delta_{h_j+1,h_{j+1}}\delta_{h'_{j}+1,h'_{j+1}}\left(+\delta_{h_{j},h_{j+1}}\delta_{h'_{j},h'_{j+1}}\right)\right].   
\end{split}
     \label{eq: FredkinTransferMatrix}
\end{equation}
The first term corresponds to spin up,  the second term to spin down and the third term to spin-$0$ (only present in the Motzkin case), with $0\le h_{j}, h'_{j} \le L$ for all $j$. From now on, we suppress the four indices of the TM and write $\mathcal{T}$. We can also write \cref{eq: FredkinTransferMatrix} diagrammatically as seen in \cref{fig: transferMatrixDefinition} (a) for the Fredkin case.
\begin{figure}[hbt!]
    \centering
    \scalebox{0.77}{
    \begin{tikzpicture}
        \begin{scope}[scale = 1, shift = {(0, 0)}]
            \node at (-1.3, -0.25) {$\mathcal{T}=$};
            \node at (-0.5, 0.75) {$h_{j}$};
            \node at (1.6, 0.75) {$h_{j+1}$};
            \drawAMPSup{0, 0, 0}{1}{A^{(s_j)}}
            \node at (0.75, -0.4) {$s_{j}$};
            \drawAconjugateMPS{0, -1.75, 0}{1}{\bar{A}^{(s_j)}}
            \node at (-0.5, -1) {$h'_{j}$};
            \node at (1.6, -1) {$h'_{j+1}$};
        \end{scope}
        \begin{scope}[scale = 1, shift = {(2.3, 0)}]
            \node at (0, -0.4) {$=$};
        \end{scope}
        \begin{scope}[scale = 1, shift = {(3.5, 0)}]
            \node at (-0.4, 0.75) {$h$};
            \node at (1.6, 0.75) {$h +1$};
            \drawAMPSup{0, 0, 0}{1}{A^{(+)}}
            \node[scale = 0.9] at (0.75, -0.4) {$+$};
            \drawAconjugateMPS{0, -1.75, 0}{1}{\bar{A}^{(+)}}
            \node at (-0.4, -1) {$h'$};
            \node at (1.6, -1) {$h'+1$};
        \end{scope}
        \begin{scope}[scale = 1, shift = {(5.8, 0)}]
            \node at (0, -0.4) {$+$};
        \end{scope}
        \begin{scope}[scale = 1, shift = {(7, 0)}]
            \node at (-0.5, 0.75) {$h+1$};
            \node at (1.4, 0.75) {$h$};
            \drawAMPSup{0, 0, 0}{1}{A^{(-)}}
            \node[scale = 0.9] at (0.75, -0.4) {$-$};
            \drawAconjugateMPS{0, -1.75, 0}{1}{\bar{A}^{(-)}}
            \node at (-0.5, -1) {$h'+1$};
            \node at (1.4, -1) {$h'$};
            
        \end{scope}
        \begin{scope}[shift = {(12, 0)}]
                    \begin{scope}[shift = {(0, 0)}, scale = 0.7]
            \node[scale = 0.8] at (0, 0) {$\mathcal{T} = \begin{pmatrix}
            0 & \textcolor{blue}{q} & 0   & 0   & 0   & 0   & 0 & 0 & 0\\
            \textcolor{blue}{q} & 0 & \textcolor{blue}{q^{3}} & 0   & 0   & 0   & 0 & 0 & 0 \\
            0   & \textcolor{blue}{q^{3}} & 0 & 0 & 0   & 0   & 0 & 0 & 0\\
            0   & 0   & 0 & 0 & \textcolor{red}{q^{2}} & 0   & 0 & 0 &0\\
            0   & 0   & 0   & \textcolor{red}{q^{2}} & 0 & 0 & 0 & 0 & 0\\
            0   & 0   & 0   & 0   & 0 & 0 & \textcolor{green}{q^{2}} & 0 &0 \\
            0   & 0   & 0   & 0   & 0   & \textcolor{green}{q^{2}} & 0 & 0 &0\\
            0   & 0   & 0   & 0   & 0   & 0   & 0   & 0 & 0\\
            0   & 0   & 0   & 0   & 0   & 0   & 0   & 0 & 0
        \end{pmatrix}$};

    \end{scope}
    \begin{scope}[shift = {(-0.5, -0.3)}, scale = 0.8]
        \node[scale = 0.56] at (-1.6, -2) {$\textcolor{blue}{00}$};
        \node[scale = 0.56] at (-1, -2) {$\textcolor{blue}{11}$};
        \node[scale = 0.56] at (-0.3, -2) {$\textcolor{blue}{22}$};
        \node[scale = 0.56] at (0.4, -2) {$\textcolor{red}{01}$};
        \node[scale = 0.56] at (1.1, -2) {$\textcolor{red}{12}$};
        \node[scale = 0.56] at (1.8, -2) {$\textcolor{green}{10}$};
        \node[scale = 0.56] at (2.5, -2) {$\textcolor{green}{21}$};
        \node[scale = 0.56] at (3.1, -2) {$02$};
        \node[scale = 0.56] at (3.7, -2) {$20$};
    \end{scope}
    \begin{scope}[shift = {(0.8, 0.3)}, scale = 0.8]

        \node[scale = 0.56] at (2.6, 1.55) {\textcolor{blue}{$00$}};
        \node[scale = 0.56] at (2.6, 1.05) {\textcolor{blue}{$11$}};
        \node[scale = 0.56] at (2.6, 0.6) {\textcolor{blue}{$22$}};
        \node[scale = 0.56] at (2.6, 0.12) {\textcolor{red}{$01$}};
        \node[scale = 0.56] at (2.6, -0.35) {\textcolor{red}{$12$}};
        \node[scale = 0.56] at (2.6, -0.83) {\textcolor{green}{$10$}};
        \node[scale = 0.56] at (2.6, -1.3) {\textcolor{green}{$21$}};
        \node[scale = 0.56] at (2.6, -1.77) {$02$};
        \node[scale = 0.56] at (2.6, -2.28) {$20$};
        
    \end{scope}
        \begin{scope}[shift = {(3.5, 1.1)}, scale = 0.7]
            \draw[thick, blue] (0, 0) circle(0.3);
            \node[scale = 0.7] at (0, 0) {$\textcolor{blue}{00}$};
            \draw[thick, blue] (0.3, 0) -- (0.6, 0);
            \draw[thick, blue] (0.9, 0) circle(0.3);
            \node[scale = 0.7] at (0.9, 0) {$\textcolor{blue}{11}$};
            \draw[thick, blue] (1.2, 0) -- (1.5, 0);
            \draw[thick, blue] (1.8, 0) circle(0.3);
            \node[scale = 0.7] at (1.8, 0) {$\textcolor{blue}{22}$};
        \end{scope}
        \begin{scope}[shift = {(3.5, 0.2)}, scale = 0.7]
            \draw[thick, red] (0, 0) circle(0.3);
            \node[scale = 0.7] at (0, 0) {$\textcolor{red}{01}$};
            \draw[thick, red] (0.3, 0) -- (0.6, 0);
            \draw[thick, red] (0.9, 0) circle(0.3);
            \node[scale = 0.7] at (0.9, 0) {$\textcolor{red}{12}$};

        \end{scope}

        \begin{scope}[shift = {(3.5, -0.5)}, scale = 0.7]
            \draw[thick, green] (0, 0) circle(0.3);
            \node[scale = 0.7] at (0, 0) {$\textcolor{green}{10}$};
            \draw[thick, green] (0.3, 0) -- (0.6, 0);
            \draw[thick, green] (0.9, 0) circle(0.3);
            \node[scale = 0.7] at (0.9, 0) {$\textcolor{green}{21}$};

        \end{scope}
        \begin{scope}[shift = {(3.5, -1.3)}, scale = 0.7]
            \draw[thick] (0, 0) circle(0.3);
            \node[scale = 0.7] at (0, 0) {$20$};
            \draw[thick] (0.9, 0) circle(0.3);
            \node[scale = 0.7] at (0.9, 0) {$02$};
        \end{scope}
        \end{scope}
    \begin{scope}[shift={(4,-2.5)}]
      \node at (0,0) {(a)};
      \node at (8.5,0) {(b)};
    \end{scope}
    \end{tikzpicture}
    }
    \caption{(a) Diagrammatic depiction of the Fredkin TM (left) and its nonzero elements (right), where $0\le h, h' \le L-1$. (b) The block diagonal Fredkin TM for $L = 2$, and the corresponding disjoint path graphs. The two numbers marking the rows (resp.~columns) denote the values of the $h_{j}h'_{j}$ (resp.~$h_{j+1}h'_{j+1}$) indices and the nodes in the graph are labeled by the values of the indices of the TM.}
    \label{fig: transferMatrixDefinition}
\end{figure}

The indices $h_j$ and $h_j'$ can in general take on any value between $0$ and $L$, but the non-vanishing entries of the TM satisfy $h_{j}-h_{j+1}=h'_{j}-h_{j+1}'=\pm 1$, or additionally 0 for Motzkin. This means that the TM is block diagonal. Furthermore, each block is a symmetric matrix, due to the parity between the physical spin configurations being contracted. The TM thus corresponds to the adjacency matrix for a disjoint union of path graphs. The adjacency matrix and corresponding path graph is seen for the $L = 2$ system for the Fredkin chain in \cref{fig: transferMatrixDefinition} (b). Note that the two numbers marking the rows (resp.~columns) denote the values of the $h_{j}h'_{j}$ (resp.~$h_{j+1}h'_{j+1}$) indices. In the Motzkin TM one also has nonzero entries on the diagonal corresponding to flat moves in the Motzkin walks and thus loops in the graph that connect nodes to themselves.  In general the $\mathcal{T}$ is written as
\begin{equation}
    \mathcal{T} =\bigoplus_{b=1}^{2L+1} T_{b} =
    \begin{pmatrix}
    T_{1} & 0 & \cdots & 0 \\
    0 & T_{2} & \cdots & 0 \\
    \vdots & \vdots & \ddots & \vdots \\
    0 & 0 & \cdots & T_{2L+1}
    \end{pmatrix},
    \label{eq: T_block_diag_direct sum}
\end{equation}
where $\bigoplus$ is the direct sum. Each block $b$ of dimension $d_{b}$ corresponds to one of the path graphs and the corresponding matrix $T_{b}$ is given as the adjacency matrix to that graph, with $q$-dependent weights (see \cref{fig: transferMatrixDefinition} for the $L = 2$ case). The critical exponents will be computed by evaluating expectation values. They can be computed from the TM as
\begin{equation}
\begin{split}
    &\begin{tikzpicture}
                \begin{scope}[scale = 0.6]
                \fill[white] (1, 1) circle (1pt); 
                \node[scale =1.1] at (-1, -0.8) {$\langle O_{r}\rangle=\frac{1}{C}$};
                \fill[black] (1.1, 0.2) circle (5pt);
                \drawAMPS{2, -0.3, 0}{0.7}{A^{(s_{1})}} 
                \drawAconjugateMPS{2, -2.5, 0}{0.7}{\bar{A}^{(s_{1})}}
                \fill[black] (1.1, -2) circle (5pt);
                \drawAMPS{4, -0.3, 0}{0.7}{A^{(s_{2})}} 
                \drawAconjugateMPS{4, -2.5, 0}{0.7}{\bar{A}^{(s_{2})}}
                \node at (6.25, 0.2) {$\dots$};
                \node at (6.25, -2) {$\dots$};
            \end{scope}
            \begin{scope}[scale = 0.6, shift = {(1.7, 0)}]
                \drawAMPSshort{5.75, -0.3, 0}{0.7}{A^{(s_{r})}} 
                \draw (6.25, -0.9) circle(0.35);
                \node[scale = 0.7] at (6.25, -0.9) {$O$};
                \drawAconjugateMPSshort{5.75, -2.5, 0}{0.7}{\bar{A}^{(s_{r})}} 
                \node at (8, 0.2) {$\dots$};
                \node at (8, -2) {$\dots$};
                \drawAMPS{9.2, -0.3, 0}{0.62}{A^{(s_{2L})}}
                \fill[black] (11.1, 0.2) circle (5pt);
                \drawAconjugateMPS{9.2, -2.5, 0}{0.62}{\bar{A}^{(s_{2L})}}
                \fill[black] (11.1, -2) circle (5pt);
            \end{scope}
    \end{tikzpicture}\\
         &\;\;\;\;\;\;\;\;\;= \frac{1}{C}\langle00|\mathcal{T}^{r-1}\mathcal{T}_{O}\mathcal{T}^{2L-r}|00\rangle,
\end{split}
\label{eq: ExpectationValueGeneralStart}
\end{equation}
where $C=\langle \mathrm{GS}| \mathrm{GS}\rangle$ is the normalization constant, and $\mathcal{T}_{O}$ is given as 
\begin{equation}
(\mathcal{T}_O)_{\,h_j, h_j';\,h_{j+1}, h_{j+1}'} = 
\sum_{s,s'=\pm} 
A^{(s)}_{h_j,h_{j+1}}O_{s s'} 
\bar{A}^{(s')}_{h_j',h_{j+1}'}.
\label{eq: T_o}
\end{equation}
The boundary conditions fixing the start and endpoints of the chain to zero height, are enforced by multiplication with the unit vector $|00\rangle = \begin{pmatrix}
    1&0&\dots&0
\end{pmatrix}$, where we use the same basis as the TM is written in \cref{fig: transferMatrixDefinition} (b).

Inserting the spectral decomposition of $\mathcal{T}$, found by decomposing each individual block $T_{b}$, gives
\begin{equation}
    \begin{split}
        \langle O_{r}\rangle= \frac{1}{C}\sum_{b=1}^{2L+1}\sum^{d_{b}}_{k_{1}, k_{2} = 1}\lambda^{r-1}_{k_{1}}&\lambda^{2L-r}_{k_{2}}\langle00|k^{b}_{1}\rangle\langle k^{b}_{1}|\mathcal{T}_{O}|k^{b}_{2}\rangle\langle k^{b}_{2}|00\rangle.
    \end{split}
    \label{eq: ExpectationValueGeneral2}
\end{equation}
Note that the entries of the eigenvectors $|k^{b}_{i}\rangle$ lying outside block $b$ are 0. This means that $\langle00|k^{b}\rangle = 0$ if $b$ is not the block in the upper left corner marked by two equal numbers, see \cref{fig: transferMatrixDefinition} (b). This matrix block has the largest dimension of any block, with dimension $d_{\text{max}} = L+1$. We denote this matrix by $T^{\text{F}}_{\text{max}}$ and $T^{\text{M}}_{\text{max}}$  for the Fredkin and Motzkin case respectively, and they have the following tridiagonal forms
\begin{equation}
\begin{aligned}
\big(T^{\mathrm{F}}_{\max}\big)_{jl} &=
\begin{cases}
q^{j+l-2}, & |j-l| =1, \\[4pt]
0, & \text{otherwise},
\end{cases}
&
\quad
\big(T^{\mathrm{M}}_{\max}\big)_{jl} &=
\begin{cases}
q^{\,j+l-2}, & |j-l|\le 1, \\[4pt]
0, & \text{otherwise},
\end{cases}
\end{aligned}
\label{eq:transferMatrixBlockDeformedFredkinComponents}
\end{equation} with $j, l \in[1, \dots, L+1]$.
Thus, \cref{eq: ExpectationValueGeneral2} can be rewritten only in terms of the eigendecomposition of $T_{\text{max}}$, which reads
\begin{equation}
    \begin{split}
        \langle O_{r}\rangle &=\frac{1}{C}\sum^{d_{\text{max}}}_{k_{1}, k_{2} = 1}\lambda^{r-1}_{k_{1}}\lambda^{2L-r}_{k_{2}}\langle00|k_{1}\rangle\langle k_{1}|\mathcal{T}_{O}|k_{2}\rangle\langle k_{2}|00\rangle ,\\
    &=\frac{1}{C}\sum^{d_{\text{max}}}_{k_{1}, k_{2} = 1}\lambda^{r-1}_{k_{1}}\lambda^{2L-r}_{k_{2}}c_{k_{1}}\mathcal{T}^{k_{1}k_{2}}_{O}c_{k_{2}},
    \end{split}
    \label{eq: ExpectationValueGeneralFinal}
\end{equation} where $|k_{i}\rangle$ are the eigenvectors of $T_{\text{max}}$, $c_{k_{i}} = \langle00|k_{i}\rangle = \langle k_{i}|00\rangle$ and
\begin{equation}
    \begin{split}
        \mathcal{T}^{k_{1}k_{2}}_{O} &= \langle k_{1}|\mathcal{T}_{O}|k_{2}\rangle=\sum^{d_{\text{max}}}_{j, l = 1}{}_{j}\langle k_{1}|(\mathcal{T}_{O})_{jl}|k_{1}\rangle_{l}.
    \end{split}
    \label{eq: To_k1k2_general}
\end{equation}
Here $|k_{i}\rangle_{l}$ denote the $l$'th component of the vector, again using the basis used in \cref{fig: transferMatrixDefinition} (b).
Since the components of $|k_{i}\rangle$ vanish outside subspace of the maximum block, we can restrict the sum to indices $j, l$ within the block $T_{\text{max}}$. As an immediate consequence of this, we get $\langle S^{x}_{r}\rangle = \langle S^{y}_{r}\rangle = 0$ for all $r$. This is seen by inserting the Pauli matrices $\sigma_{x}$ and $\sigma_{y}$ for $O$ in \cref{eq: T_o}. Since these Pauli matrices act as `bit flips', we find that all entries located by two indices denoted by two equal numbers, that is the indices within $T_{\text{max}}$, are zero. Thus, for the GS of the deformed Fredkin spin chain we have $\langle S^{x}_{r}\rangle = \langle S^{y}_{r}\rangle = 0$ for all $r$ and for all values of $q$. This matches previous results for the $q = 1$ GS of the Motzkin and Fredkin chain \cite{10.1063/1.4977829, Dell_Anna_2019}.

\section{Critical correlation from the TM method}\label{sec:correlation}

For gapped GSs, the long distance behavior of \cref{eq: ExpectationValueGeneralFinal} is dominated by the largest two eigenvalues of the TM. When the system is gapless, as is the case of the models in question at the critical point $q=1$, the spectrum of the TM is continuous in the thermodynamic limit. Luckily, for $q = 1$, the matrices $T^{\text{F/M}}_\text{max}$ in \cref{eq:transferMatrixBlockDeformedFredkinComponents} becomes Toeplitz, and their eigensystems have closed-from expressions that allow us to convert the discrete sum in \cref{eq: ExpectationValueGeneralFinal} to an integral.

A $d_{\text{max}}$-dimensional tridiagonal Toeplitz matrix with 0 (Fredkin case) or 1 (Motzkin case) on the diagonal and 1 on its super- and sub-diagonals has eigenvalues $\lambda_k$ and eigenvectors $|k\rangle$ for $1\le k \le d_{\text{max}}$ given by
\begin{equation}
\begin{aligned}
\lambda_{k} &=
\begin{cases}
2\cos\!\left(\dfrac{k\pi}{d_{\text{max}}+1}\right), & \text{Fredkin}, \\[1.2em]
1+2\cos\!\left(\dfrac{k\pi}{d_{\text{max}}+1}\right), & \text{Motzkin},
\end{cases}
&
\quad
|k\rangle &= \sqrt{\tfrac{2}{d_{\text{max}}}}
\begin{pmatrix}
\sin\!\big(\tfrac{k\pi}{d_{\text{max}}+1}\big) \\[6pt]
\sin\!\big(\tfrac{2k\pi}{d_{\text{max}}+1}\big) \\[4pt]
\vdots \\[4pt]
\sin\!\big(\tfrac{d_{\text{max}} k\pi}{d_{\text{max}}+1}\big)
\end{pmatrix}.
\end{aligned}
\label{eq: eigenToeplitz}
\end{equation}

Clearly, $T^\mathrm{F/M}_\text{max}$ is gapless for $q = 1$ in the large $L$ limit, as $d_{\text{max}} = L+1$. 
As discussed in \Cref{sec:transfermatrix}, to estimate the critical exponent $\eta$ normally associated with the connected correlation function in the critical phase, we will compute the one-point function $\langle S^{z}_{r}\rangle$. To that end, we insert the matrix representation of $S^{z}$, the spin-$\frac{1}{2}$ (resp.~spin-$1$) version for the Fredkin chain (resp.~Motzkin chain),  for $O$ in \cref{eq: ExpectationValueGeneralFinal}. The matrix element $\mathcal{T}^{k_{1}k_{1}}_{S^{z}}$ as defined by \cref{eq: To_k1k2_general} is shown in \Cref{sec:exp_val_derivation} to be of the form
\begin{equation}
    \begin{split}
        \mathcal{T}^{k_{1}k_{2}}_{S^{z}} &=\frac{S}{L}\left(1 -(-1)^{k_{1} + k_{2}}\right)\sin{\left(\frac{\pi k_{2}}{L}\right)}\left[\cot{\left(\frac{\pi(k_{1} + k_{2})}{2L}\right)} + \cot{\left(\frac{\pi(k_{1} - k_{2})}{2L}\right)}\right]
    \end{split},
    \label{eq: OperatorOverlapSz_final}
\end{equation}where $S = \frac{1}{2}$ (resp.~$S = 1$) for the Fredkin chain (resp.~Motzkin chain). Inserting it along with the eigenvalues and vectors in \cref{eq: eigenToeplitz}, into \cref{eq: ExpectationValueGeneralFinal} gives 
\begin{equation}
    \begin{split}
        \langle S^{z}_{r}\rangle = \frac{2S}{CL^{2}}\sum^{L}_{k_{1}, k_{2} = 1}&\left(1 -(-1)^{k_{1} + k_{2}}\right)\lambda^{r-1}_{k_{1}}\lambda^{2L-r}_{k_{2}}\sin{\left(\frac{\pi k_{1}}{L}\right)}\sin^{2}{\left(\frac{\pi k_{2}}{L}\right)}\\
        &\times\left[\cot{\left(\frac{\pi(k_{1} + k_{2})}{2L}\right)} + \cot{\left(\frac{\pi(k_{1} - k_{2})}{2L}\right)}\right],
    \end{split}
    \label{eq: Sz_expect_start}
\end{equation}where we have identified $d_{\text{max}} +1$ with $L$ since we are interested in the large $L$ limit. To make progress, we approximate the sums over $k_{1}$ and $k_{2}$ as integrals over $x = \frac{\pi k_{1}}{L}$ and $y = \frac{\pi k_{2}}{L}$, using the standard method for replacing sums with integrals valid for large $L$
\begin{equation}
    \sum^{L}_{k_{1}, k_{2} = 1}f\left(\frac{\pi k_{1}}{L}, \frac{\pi k_{2}}{L}\right) \rightarrow \frac{L^2}{\pi^2}\iint^{\pi}_{0}dxdyf(x, y).
    \label{eq: Riemann_Sum_approx}
\end{equation}
Making the substitution \cref{eq: Riemann_Sum_approx}, gives 
\begin{equation}
    \begin{split}
        \langle S^{z}_{r}\rangle =\frac{2S}{C\pi^{2}}\iint^{\pi}_{0} dx dy&\lambda^{r-1}{(x)\lambda^{2L-r}(y)}\sin{x}\sin^{2}{y}\left[\cot{\left(\frac{x+y}{2}\right)} + \cot{\left(\frac{x-y}{2}\right)}\right].
    \end{split}
    \label{eq: Sz_expect_integral}
\end{equation}
In the above equation, the fast oscillating factor $(1-(-1)^{k_{1}+k_{2}})$ in \cref{eq: Sz_expect_start}, has been set to its average value of $1$. For this factor, the value changes between $0$ and $2$ when $k_{1} + k_{2}$ changes from even to odd. This corresponds to very small changes in $x$ and $y$, under which the rest of the integrand changes slowly. In the regions where the cotangent factor diverges, the divergence is either canceled by the sine or cosine factors. Ultimately, the validity of the approximations made in this section will be checked against the MPS numerical results in presented at the end. 

For $q = 1$, the normalization constant $C$ in \cref{eq: Sz_expect_integral} is simply the number of Dyck walks (resp.~Motzkin walks) on $2L$ steps for the Fredkin chain (resp.~Motzkin chain), see \Cref{sec:FredkinMotzkinSpinChain}. Both numbers have closed form expression in terms of $L$, and for large $L$ the numbers can be approximated using the Stirling approximation and \cref{eq: Sz_expect_integral} becomes 
\begin{equation}
    \langle S^{z}_{r}\rangle \approx \frac{KL^{\frac{3}{2}}}{\lambda_{*}\pi^{{\frac{3}{2}}}}\!\iint^{\pi}_{0} \!\!dx dy \left(\frac{\lambda(x)}{\lambda_{*}}\right)^{\!\!r-1}\!\!\left(\frac{\lambda(y)}{\lambda_{*}}\right)^{\!\!2L-r}\!\!\!\!\!\sin{x}\sin^{2}{y}\left[\cot{\left(\frac{x+y}{2}\right)} + \cot{\left(\frac{x-y}{2}\right)}\right],
    \label{eq: Sz_expect_final_integral}
\end{equation}with $K = 1$ (resp.~$K = \frac{8\sqrt{6}}{9}$) for the Fredkin (resp.~Motzkin chain). $\lambda_{*}$ is the maximum eigenvalue as $L\rightarrow\infty$, which is $\lambda_{*} = 2$ (resp.~$\lambda_{*} = 3$) for the Fredkin case (resp.~Motzkin case).

To evaluate the double integral, we assume that $1\ll r\ll L$. This means that $r$ is on the left side, far away from the halfway point. In this regime the main contribution to the integrand comes from the region where $x$ and $y$ are close to either $0$ or $\pi$. This can be understood as follows. Clearly, the main contributions to the integral come from the vicinity of $y = 0$ or $y = \pi$ due to the $\cos^{2L-r}y$-factor, as $(2L-r)\gg1$. At the same time, for small $r$, the $\cos^{r}x$ does not constrain $x$ to be close to $0$ or $\pi$ for the integrand in \cref{eq: Sz_expect_final_integral} to be non-vanishing. However, the $\cot(\frac{x\pm y}{2})$ factors will diverge when $x \rightarrow 0$ or $x \rightarrow\pi$, for $y\rightarrow0$ or $y\rightarrow\pi$ respectively. Thus, the main contributions to the integral come from $x = 0$ and $x =\pi$, also for small $r$. This allows for a saddle-point approximation of the double integral, which is carried out in \Cref{sec:IntegralEval}, giving 
\begin{equation}
    \langle S^{z}_{r}\rangle \approx \begin{cases} \displaystyle \frac{1}{\sqrt{2\pi(r-1)}} +\frac{(-1)^{r+1}}{4\sqrt{\pi}\left(2(r-1)\right)^{3/2}}, & \text{for Fredkin}, \\[0.56cm] \displaystyle \frac{4}{\sqrt{3\pi(r-1)}}, & \text{for Motzkin}. \end{cases}
\label{eq: Sz_expectation_expression}
\end{equation}
This shows that $\langle S^{z}_{r}\rangle \sim1/\sqrt{r}$ for both the Fredkin and Motzkin chain for $1\ll r\ll L$, which agrees with the combinatorial results of Ref.~\cite{10.1063/1.4977829, Menon:2024vic} for the Motzkin chain, Ref.~\cite{Udagawa_2017,PhysRevB.99.054436} for the Fredkin chain, and the field theoretic results in the scaling limit of Ref. \cite{Chen_2017,mykland2025highlyentangled2dground}. The corresponding critical exponent is $\eta=\frac{3}{2}$.\\

\begin{figure*}[t]
  \centering
  \begin{subfigure}[t]{0.48\linewidth}
    \includegraphics[width=\linewidth]{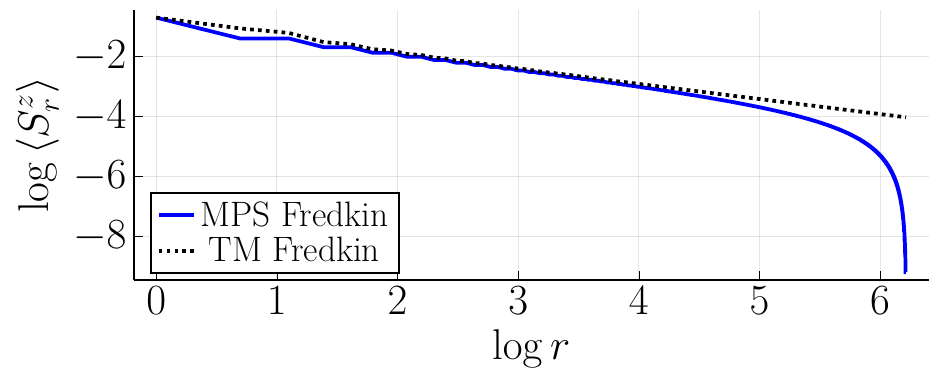}
    \caption{}
  \end{subfigure}
  \hfill
  \begin{subfigure}[t]{0.48\linewidth}
    \includegraphics[width=\linewidth]{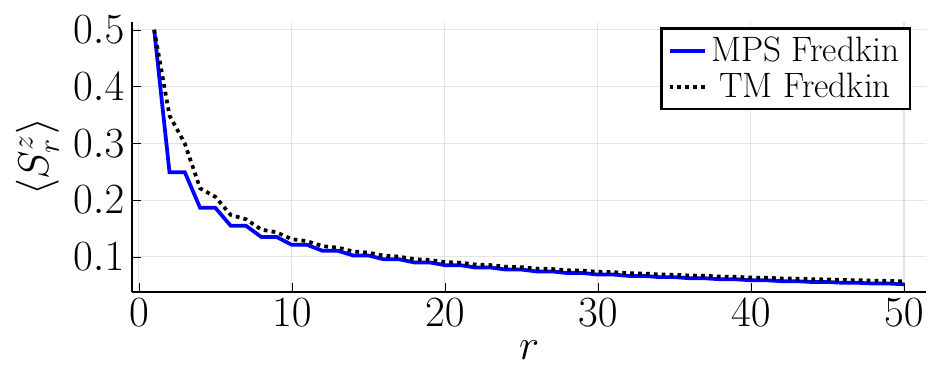}
    \caption{}
  \end{subfigure}

  \vspace{1em}

  \begin{subfigure}[t]{0.48\linewidth}
    \includegraphics[width=\linewidth]{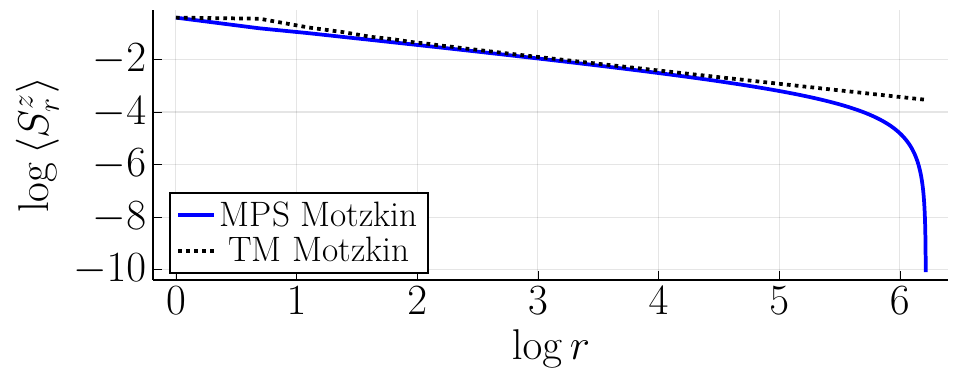}
    \caption{}
  \end{subfigure}
  \hfill
  \begin{subfigure}[t]{0.48\linewidth}
    \includegraphics[width=\linewidth]{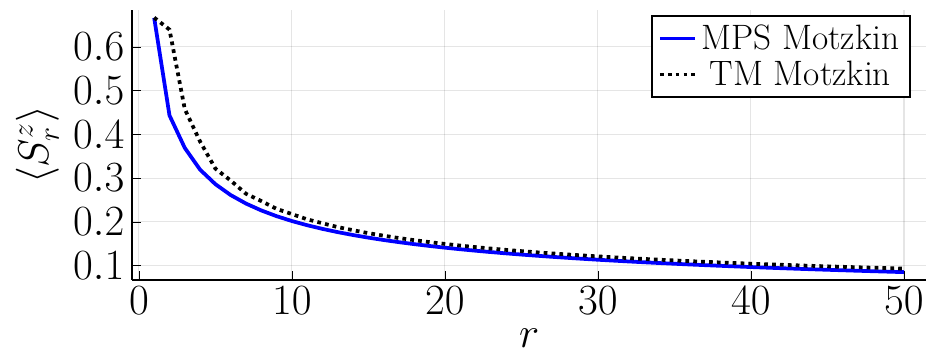}
    \caption{}
  \end{subfigure}

  \caption{Comparison of $\langle S^{z}_{r}\rangle$ obtained from the MPS ($L=500$) and from the analytic result in \cref{eq: Sz_expectation_expression} obtained from the TM for the Fredkin and Motzkin chains. Panels (a), (b) correspond to the Fredkin chain, and panels (c), (d) to the Motzkin chain. Note that panels (a) and (c) are log-log plots.}
  \label{fig:Sz_numeric_vs_analytic_Fredkin_Motzkin}
\end{figure*}

Finally, the predictions of \cref{eq: Sz_expectation_expression} are compared with the numerical results from MPS calculations in \cref{fig:Sz_numeric_vs_analytic_Fredkin_Motzkin}. The results are in agreement for $1\ll r\ll L$, when our approximations are valid. \cref{fig:Sz_numeric_vs_analytic_Fredkin_Motzkin} (b) shows that \cref{eq: Sz_expectation_expression} also captures the oscillations near the boundary. This effect vanishes as $\sim r^{-3/2}$ as seen from \cref{eq: Sz_expectation_expression}. 

\section{Critical exponents from dual zero-dimensional Hamiltonian}\label{sec:0D}

Although the derivation of the critical exponent $\eta$ in the last section and \Cref{sec:IntegralEval} seems technical, the idea behind is really quite simple. This can be made more transparent by introducing a 0D dual Hamiltonian to the TM describing the GS superposition. Define the dual Hamiltonian $\mathcal{H}$ by
\begin{equation}
    \mathcal{T}=e^{-\mathcal{H}},
\end{equation}such that its eigenvalues $\{\varepsilon_i\}$ are related to the eigenvalues of the TM $\{\lambda_i\}$ by
\begin{equation}
    \varepsilon_i=-\log \lambda_i.
\label{eq:dual_Hamiltonian_eig_val}
\end{equation}It is well known that in the case of a gapped GS, the correlation function asymptotically decays geometrically as 
\begin{equation}
    \langle O_1O_r\rangle\sim \left|\frac{\lambda_2}{\lambda_1}\right|^r=e^{-(\varepsilon_2-\varepsilon_1)r}
\label{eq:gapped_correlation}
\end{equation}where $\lambda_{1,2}$ are the two largest eigenvalues of $\mathcal{T}$.

For gapless GSs, we have a continuous spectrum for both $\mathcal{T}$ and $\mathcal{H}$, which can be expressed as functions $\lambda(x)$ and $\varepsilon(x)$ of some parameter $x\in [0,a]$ with $|\lambda(0)|$ (resp.~$\varepsilon(0)$) being the maximum (resp.~minimum). The spectra may not necessarily be continuous, but it must be for some range in the neighborhood of $x=0$ for the spectrum to be gapless. So the boundary $a$ is chosen accordingly as to make $\lambda(x)$ and $\epsilon(x)$ continuous on $[0, a]$. With these notations, the gapless version of \cref{eq:gapped_correlation} becomes 
\begin{equation}
    \langle O_1O_r\rangle\sim \int_0^a dx \left|\frac{\lambda(x)}{\lambda(0)}\right|^r= \int_0^a dx e^{-\big(\varepsilon(x)-\varepsilon(0)\big)r}, \label{eq:scaling}
\end{equation}which is the essence of the computation in \Cref{sec:IntegralEval}.\footnote{Notice that the TM may very well have non-positive eigenvalues, which is the case with the Motzkin and Fredkin GSs, so their logarithm is not necessarily unique, meaning that the mapping between $\{\varepsilon_i\}$ and $\{\lambda_i\}$ is not one-to-one. But these details affect the correlator only by an multiplicative factor, not its decay behavior, as demonstrated in \cref{sec:IntegralEval}.} The model in question for \Cref{sec:IntegralEval} has $\lambda(x)=\cos(x)$, so $\varepsilon(x)\approx \varepsilon(0)+ \frac{\varepsilon''(0)}{2} x^2$, which happens to be compatible with the approximation of evaluating the integral with Laplace's method. But in general, the spectrum of the dual Hamiltonian near its maximum can behave as $\varepsilon(x)-\varepsilon(0)\approx \alpha x^\frac{1}{\eta-1}$. Then, by rescaling the integration variable, the integral in \cref{eq:scaling} evaluates to
\begin{equation}
    \langle O_1O_r\rangle\sim \int_0^\infty dx e^{-\alpha x^\frac{1}{\eta-1}r}=r^{1-\eta}\int_0^\infty du e^{-\alpha u^\frac{1}{\eta-1}} \propto r^{1-\eta},
\end{equation}where we have extended the integral to infinity without introducing much error. Thus, we have reduced the problem of finding critical exponent $\eta$ generically to diagonalizing the dual 0D Hamiltonian of the 1D TM. The dual 0D Hamiltonian for the Fredkin GS at the critical point $q=1$ is equivalent to a homogeneous 1D tight binding model in the height space, with well-known quadratic dispersion, leading to the value $\eta=3/2$. But when the hopping constant in the equivalent tight binding model is $q$-deformed, the diagonalization of the dual Hamiltonian becomes non-trivial. This will be done in the next section.

\section{RG analysis of \texorpdfstring{$q$}{q}-deformation}\label{sec:RG}

At $q\neq1$ the matrices $T_{\text{max}}$ in \cref{eq:transferMatrixBlockDeformedFredkinComponents} are no longer Toeplitz, and their eigenvalues and eigenvectors do not have closed form expressions for arbitrary $L$. But the duality (for some orthogonal matrix $S$ that inverts the column and row indices) 
\begin{equation}
    q^{2L}T(1/q)= S T(q) S^{-1}, 
\end{equation}implies that the ratio between the two largest magnitudes of their eigenvalues for $T(q)$ and $T(1/q)$ is the same. Therefore we have $\nu_\pm=\nu$, with $\nu_\pm$ defined respectively by $\xi\propto (\mp \tau)^{-\nu_\pm}$ for the ordered and disordered phases, where $\tau=-\log q$ is the `reduced temperature'. The explicit value can be determined from the scaling dimension of $S^z$ obtained in \cref{eq: Sz_expectation_expression} using RG analysis.

An expression for the free energy of the ensemble of configurations in the GS has been derived for the scaling limit in Ref.~\cite{Zhang2024quantumlozenge}. The 1D lattice version can be written as
\begin{equation}
    F_L(\sigma,\tau)=\sum_{r=1}^L \left(\sigma (S_r^z)^2 +\tau\sum_{j=1}^r S_j^z\right)=\sum_{r=1}^L \left(\sigma (S_r^z)^2 +\tau (L-r) S_r^z\right),\label{eq:freeenergy}
\end{equation}where $\sigma$ is the surface tension of the height function that correspond to the thermal parameter in the RG. Notice that the above form of the free energy is only a good approximation for $q\ne 1$. At the critical point, the hard wall condition at zero height becomes important for the scaling, which effectively makes all the spins in the chain strongly interacting with each other. Thus the non-interacting version of the free energy is only valid when the non-negativity of the height can be ignored compared to the spatially varying magnetic field $\tau(L-r)$ that vanishes at $q=1$.

After one step of coarse graining with block size $b=2$, it becomes
\begin{equation}
    F'_{L/b}(\sigma',\tau')=\sum_{r=1}^{L/b} \left(\sigma' \left(S_r^{\prime z}\right)^2 +\tau' \left(\frac{L}{b}-r\right) S_r^{\prime z}\right),
\end{equation}In order for the free energy to stay invariant, its density must satisfy $f'(\sigma',\tau')=bf(\sigma,\tau)$, which demands the RG equations
\begin{equation}
    \tau' S^{\prime z}=b^2\tau S^z, \quad \sigma' (S^{\prime z })^2= b \sigma (S^{z })^2.
\end{equation} 
As \cref{eq: Sz_expectation_expression} established the scaling dimension of the spin operator $S^{\prime z}= b^{1/2} S^z$, we must have $\tau'=b^{3/2}\tau$ and $\sigma'=\sigma$. Since the correlation length changes as $\xi'(\tau')=\xi(\tau)/b$, we have the equation
\begin{equation}
    \xi(b^{3/2}\tau)=\frac{\xi(\tau)}{b},
\end{equation}which is solved by $\xi(\tau)=\xi_0 \tau^{-2/3}$, giving $\nu=2/3$.
\begin{figure}[hbt!]
  \centering
  \begin{subfigure}[t]{0.48\linewidth}
    \includegraphics[width=\linewidth]{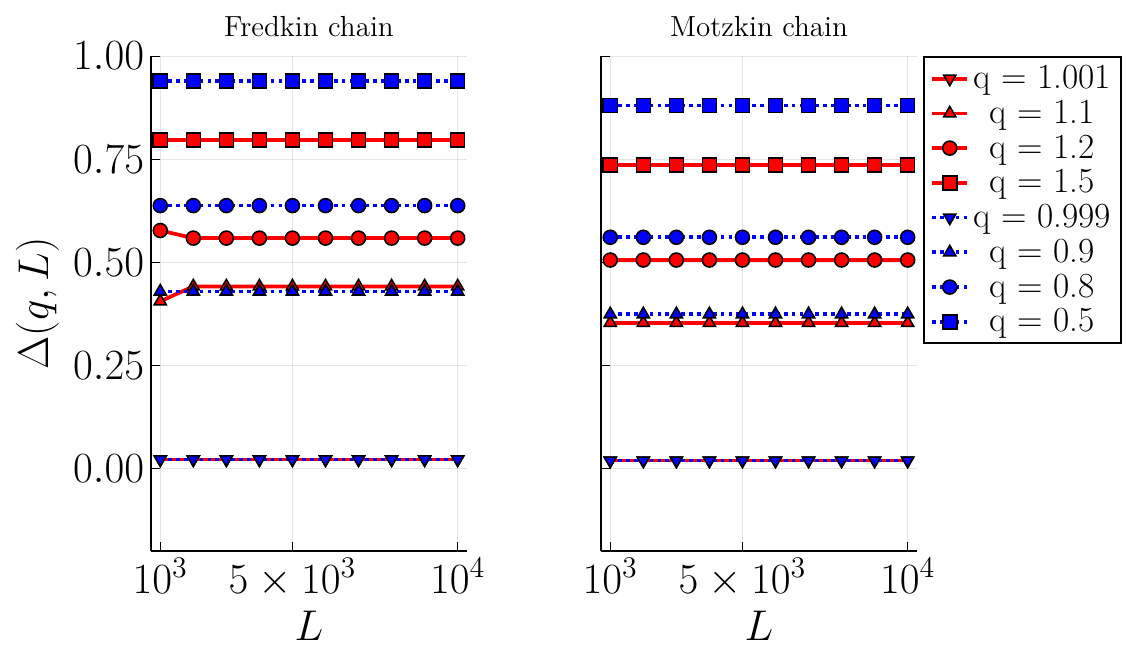}
    \caption{}
  \end{subfigure}
  \hfill
  \begin{subfigure}[t]{0.48\linewidth}
    \includegraphics[width=\linewidth]{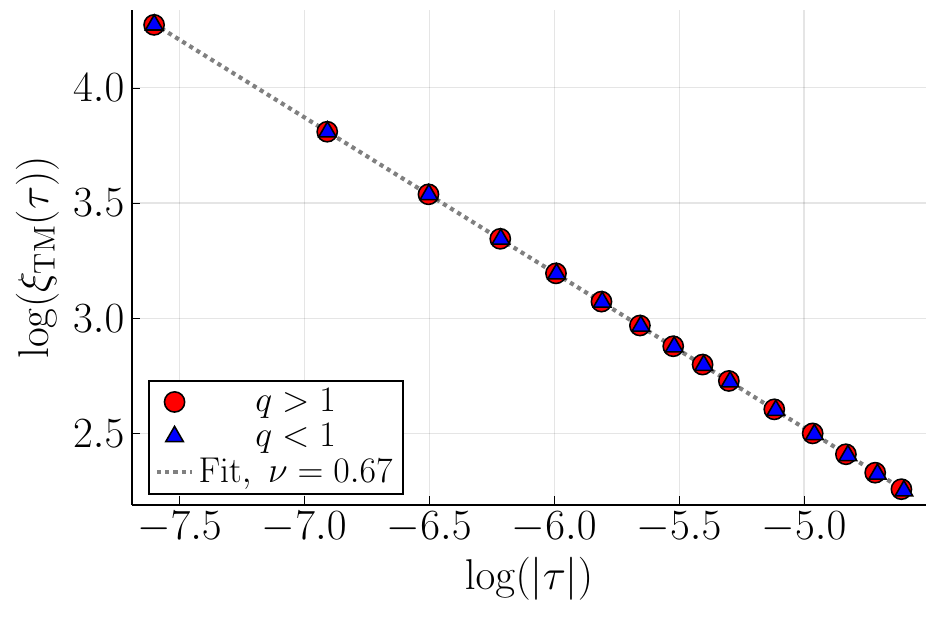}
    \caption{}
  \end{subfigure}
  \caption{(a) The eigenvalue gap $\Delta(q, L) = (\lambda_{1}-\lambda_{2})/\lambda_{1}$ as a function of the system size $L$ for various $q$s. (b) Log-log plot of the correlation length $\xi(\tau) = -1/\ln|\frac{\lambda_{2}}{\lambda_{1}}|$, where $\tau = \log q$, for various $q<1$ (marked by blue circles) and $q>1$ (marked by red triangles), with $L = 4000$. The dotted gray line indicate the fit $\xi_{\text{TM}}(\tau)\sim |\tau|^{-\nu}$. The results in (b) are for the Fredkin chain, but identical behavior is seen for the Motzkin chain.}
  \label{fig:eig_val_gaps_corr_length}
\end{figure}

This result is confirmed by the numerical diagonalization of the TMs $T_{\text{max}}$. The finite correlation length off criticality
\begin{equation}
    \xi_\mathrm{TM}(\tau)=\frac{a}{\ln \frac{\lambda_1(\tau)}{\lambda_2(\tau)}},
\end{equation}for some constant length $a$, is given by the inverse of the logarithm of the ratio between the eigenvalues with the two largest magnitudes $\lambda_{1,2}$. For $q\ne 1$, there is a finite spectral gap of $T_{\text{max}}$ in the thermodynamic limit of $d_\mathrm{max}\to \infty$, as shown in \cref{fig:eig_val_gaps_corr_length} (a). The dependence of the numerical correlation length extracted from the TM, $\xi_\mathrm{TM}(\tau)$, on the reduced temperature $\tau$, shown in log-log scale in \cref{fig:eig_val_gaps_corr_length} (b), agrees with the scaling analysis from the RG.
\begin{figure}[hbt!]
  \centering
  \begin{subfigure}[t]{0.48\linewidth}
    \includegraphics[width=\linewidth]{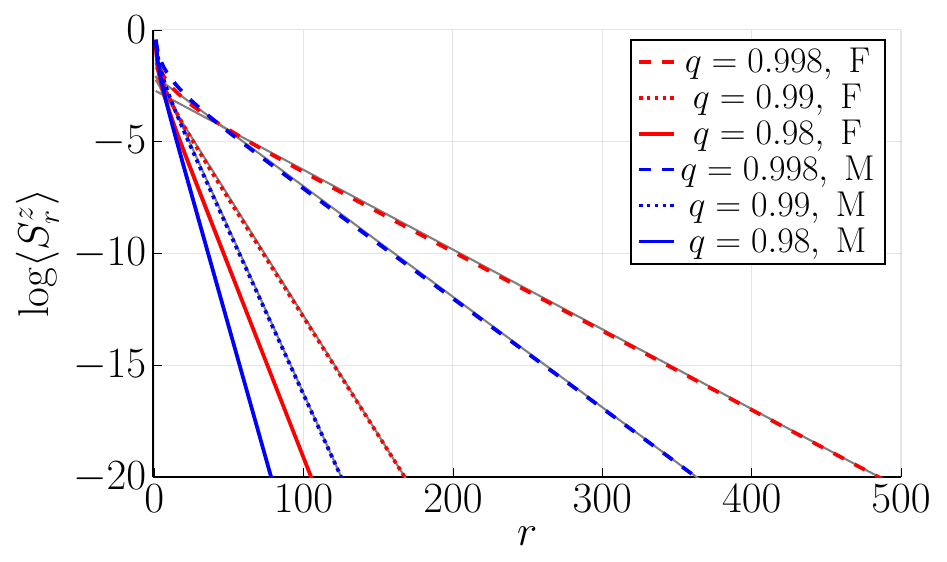}
    \caption{}
  \end{subfigure}
  \hfill
  \begin{subfigure}[t]{0.48\linewidth}
    \includegraphics[width=\linewidth]{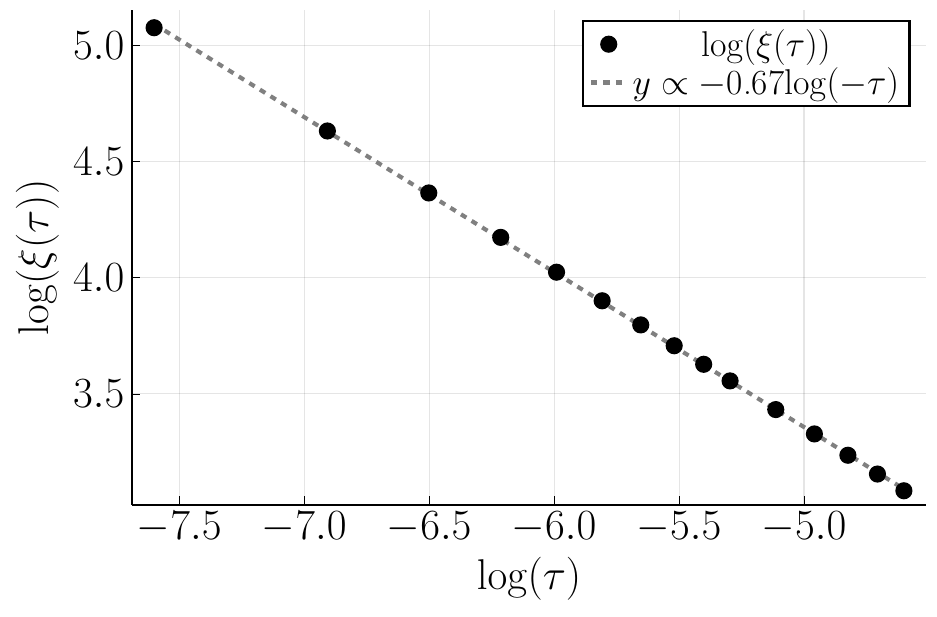}
    \caption{}
  \end{subfigure}
  \caption{(a) Plot of $\log{\langle S^{z}_{r}\rangle}$ in the GS for various $q<1$. The faint gray lines indicate the linear fits determining the correlation length $\xi(q)$. The plot is shown for the first $500$ sites with $L = 1000$ and F (resp.~M) denotes Fredkin (resp.~Motzkin). (b) Log-log plot of the correlation length $\xi(|\tau|)$ for $\tau = -\log q>0$. The dotted gray line is the line fit of slope $\approx-0.67$ and the data points are marked with black dots. Note that the results shown are for the Fredkin chain, but identical results are obtained for the Motzkin chain.}
  \label{fig:Sz_log_plots_qneq1}
\end{figure}
 
Although the correlation length is finite for both the disordered ($q<1$) and ordered ($q>1$) phases, the spatial decay of the order parameter $\langle S_r^z\rangle$ behaves quite differently in the two phases. For the disordered phase, the order parameter decays exponentially as $\langle S_r^z\rangle=e^{-r/\xi}$ for $1\ll r\ll L$, which is confirmed by the numerical results from the MPS calculation shown in \cref{fig:Sz_log_plots_qneq1}, where the $\xi(|\tau|)$ is directly extracted from the correlation function. For the ordered phase, however, due to the opposite boundary conditions, there is a domain wall with a thickness that depends on the correlation length. 
\begin{figure}[hbt!]
  \centering
  \begin{subfigure}[t]{0.48\linewidth}
    \includegraphics[width=\linewidth]{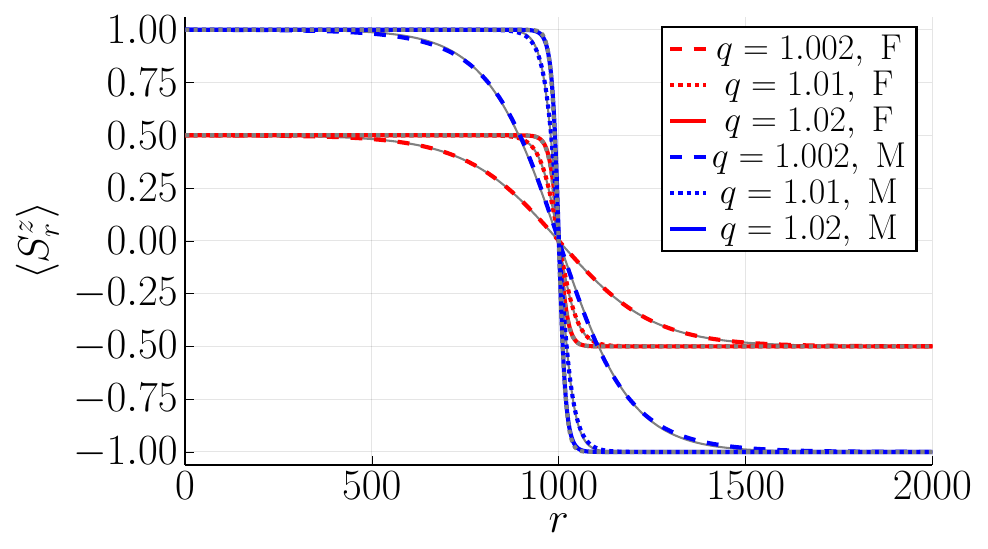}
    \caption{}
  \end{subfigure}
  \hfill
  \begin{subfigure}[t]{0.48\linewidth}
    \includegraphics[width=\linewidth]{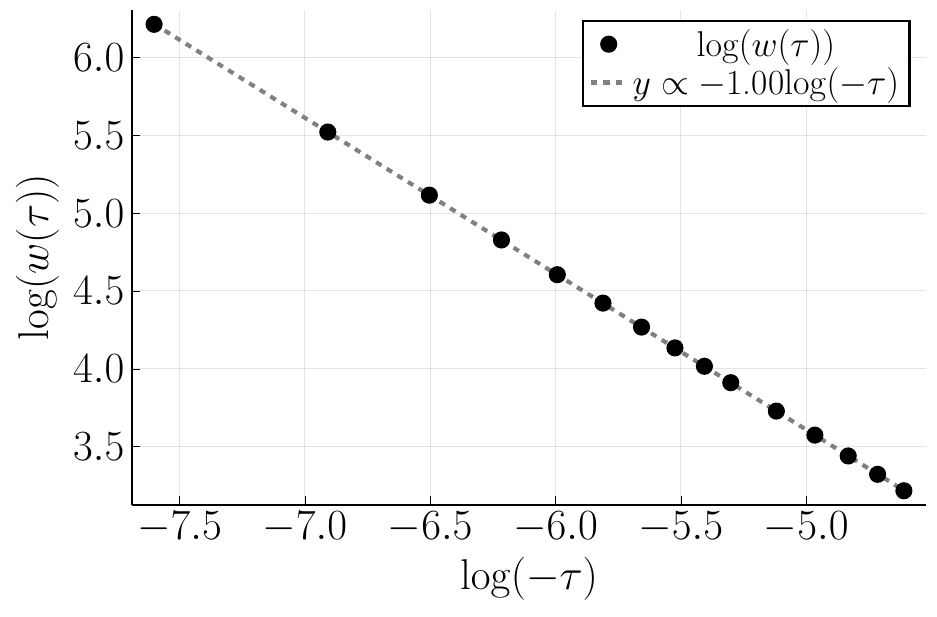}
    \caption{}
  \end{subfigure}
  \caption{(a) Plot of $\langle S^{z}_{r}\rangle$ for various $q>1$, for a system of $L = 1000$. The faint gray lines indicate the fitted Fermi–Dirac distribution (almost indistinguishable from the data) giving the domain wall width $w(q)$. Note that F (resp.~M) denotes Fredkin (resp.~Motzkin). (b) Log-log plot of the domain wall width $w(\tau)$, for $\tau = -\log q<0$. The dotted gray line is the line fit of slope $\approx-1.00$ and the data points are marked with black dots. Note that the results shown are for the Fredkin chain, but identical results are obtained for the Motzkin chain.}
  \label{fig:xi(q)_plot}
\end{figure}

Interestingly, the dependence is different from the prediction from Ginzburg-Landau theory, where the domain wall thickness is proportional to the correlation length (see \Cref{sec:MFT} for a derivation). In our case, the free energy \cref{eq:freeenergy} for the classical spin configurations in the GS superposition \cref{eq:freeenergy} is non-interacting. So the partition function decomposes into a product of the partition function at each individual lattice site. The expectation value of the spin $\langle S_r^{z}\rangle$ can hence be calculated as
\begin{equation}
    \langle S_r^{z}\rangle= S\frac{e^{\tau (r-L)  }-e^{-\tau (r-L) }}{e^{\tau (r-L)  }+e^{-\tau (r-L)}}=S\tanh\left(\tau (r-L) \right),
\end{equation}where $S =\frac{1}{2}$ (resp. $S = 1$) for the Fredkin (resp. Motzkin) chain. This novel scaling relation $w \sim \xi^{3/2}$ between the domain wall thickness $w\propto(-\tau)^{-1}$ and the correlation length $\xi\propto (-\tau)^{-2/3}$ is also confirmed by numerical results from MPS, as shown in \cref{fig:xi(q)_plot} (b). The hyperbolic tangent profile of the domain wall was first observed from numerical data in Ref.~\cite{Udagawa_2017}, but here we gave a theoretical derivation.

\section{Conclusion}\label{sec:conclusion}

Much like in classical statistical mechanics, where TM and RG are the two non-perturbative approaches for analytical solutions, MPS and PEPS on the one hand, and MERA on the other, have emerged as the two main tensor-network methods capable of delivering exact or
controlled results for quantum many-body systems. These approaches typically apply to rather different classes of models: MPS/PEPS exploit translational invariance on regular lattices, whereas MERA is naturally adapted to hierarchical structures and scale-invariant states. In this work, we have shown that the Motzkin and Fredkin GSs constitute a remarkable setting in which these two paradigms can be unified. Their random-walk structure provides an exact MPS description, while their long-range entanglement and self-similarity admit a MERA-like hierarchical TN representation. 

A central technical contribution of this paper is to extend TM methods to genuinely critical GSs. TM techniques are well established for demonstrating exponential decay of correlations and computing correlation lengths in non-critical systems. However, to the best of our knowledge, there have been no prior examples where TM is used directly at a quantum critical point to extract power-law correlation decay, precisely because the largest TM eigenvalues become degenerate and the associated correlation length diverges.
Here, starting from the exact random-walk/MPS representations of the Motzkin and Fredkin ground states, we constructed their TM and showed explicitly that the spin expectation value decays algebraically at criticality. From the TM spectrum and its continuum approximation, we determined the critical exponent $\eta$ for the spin correlations. Moreover, we interpreted the $q$-deformation away from criticality as a perturbation of the TM at the critical point, which splits the eigenvalue degeneracy. The resulting scaling of the correlation length yields a universal value of the critical exponent $\nu$ that is independent of the microscopic details of the TM.

Traditionally, critical exponents are computed via RG analyses, either field-theoretic or lattice-based. As Vidal has emphasized \cite{PhysRevLett.99.220405}, the sequence of coarse-grained configurations obtained at different layers of a MERA, or of the pentagonal TN constructed in this work, can be interpreted as an RG flow of ground states. In our TM-based analysis, we introduced an effective `temperature' $\tau = - \log q$ controlling the $q$-deformation and showed that it has an RG eigenvalue $y_\tau = 3/2$. At present, however, it is not clear how to recover this eigenvalue purely from the RG transformation encoded in the hierarchical TN itself, without referring to the TM dual description. A natural future direction is therefore to implement explicitly the RG step defined by the coarse-graining procedure in the pentagon TN, and to compute the critical exponents directly from the tensor-network RG.

Beyond these methodological insights, our results highlight the Motzkin and Fredkin chains, together with their $q$-deformations, as frustration-free spin models that realize an unconventional quantum phase transition from a disordered to an ordered phase under domain-wall boundary conditions. The bulk ordering of the spins can be interpreted as arising from their correlation with the nearer boundary spin, which is subject to an effective boundary magnetic field; related boundary-induced ordering mechanisms have been discussed in other spin chains \cite{Affleck_1998} and warrant a more systematic investigation in the present context. Using the TM constructed from the uniform MPS, we extracted the power-law decay of spin correlations at criticality and obtained the exponent $\eta$. Treating the $q$-deformation as a perturbation of the critical TM, we determined the correlation-length exponent $\nu$ and uncovered a duality between the ordered and disordered phases that is not manifest in the original Motzkin- or Dyck-path representations. These analytical exponents were corroborated by numerical MPS simulations and direct TM diagonalization.

On the tensor-network side, we showed that both the Motzkin and Fredkin ground states admit a uniform TN representation whose unit cell is effectively a rank-5 tensor. We analyzed how different decompositions of this rank-5 tensor into rank-3 and rank-4 building blocks relate the exact holographic Motzkin TN to MERA-like constructions, thereby illustrating a trade-off between uniformity and informativeness in tensor representations of the same long-range-entangled state. This suggests that rank five is the minimal tensor rank at which different TNs for one-dimensional critical ground states can share a common geometric backbone.

Taken together, our results demonstrate that frustration-free, highly entangled spin chains provide analytically tractable laboratories for quantum critical phenomena. They bridge combinatorial path representations, TM techniques, and field theoretic ideas, and show that exact critical exponents $(\eta,\nu)$ can be obtained in models whose ground states are non-Gaussian and strongly entangled.
We expect that the interplay of TM, classical-quantum duality, and TN structure uncovered here will offer a useful framework for future studies of quantum criticality, including higher-dimensional generalizations and the search for exact MERA-type representations of gapless ground states.

\section*{Acknowledgements}
Zhao Zhang thanks Long Liang and Yuan Miao for discussions.

\paragraph{Author contributions}
OM conceived the idea and performed the numerical calculations. OM and ZZ conducted the analytical part of the research and wrote the paper together.

\begin{appendix}
\numberwithin{equation}{section}
\section{Pentagonal TN for the Motzkin and Fredkin GSs}\label{sec:penta}

\begin{figure*}[ht]
    \centering
    \includegraphics[width=0.9\textwidth]{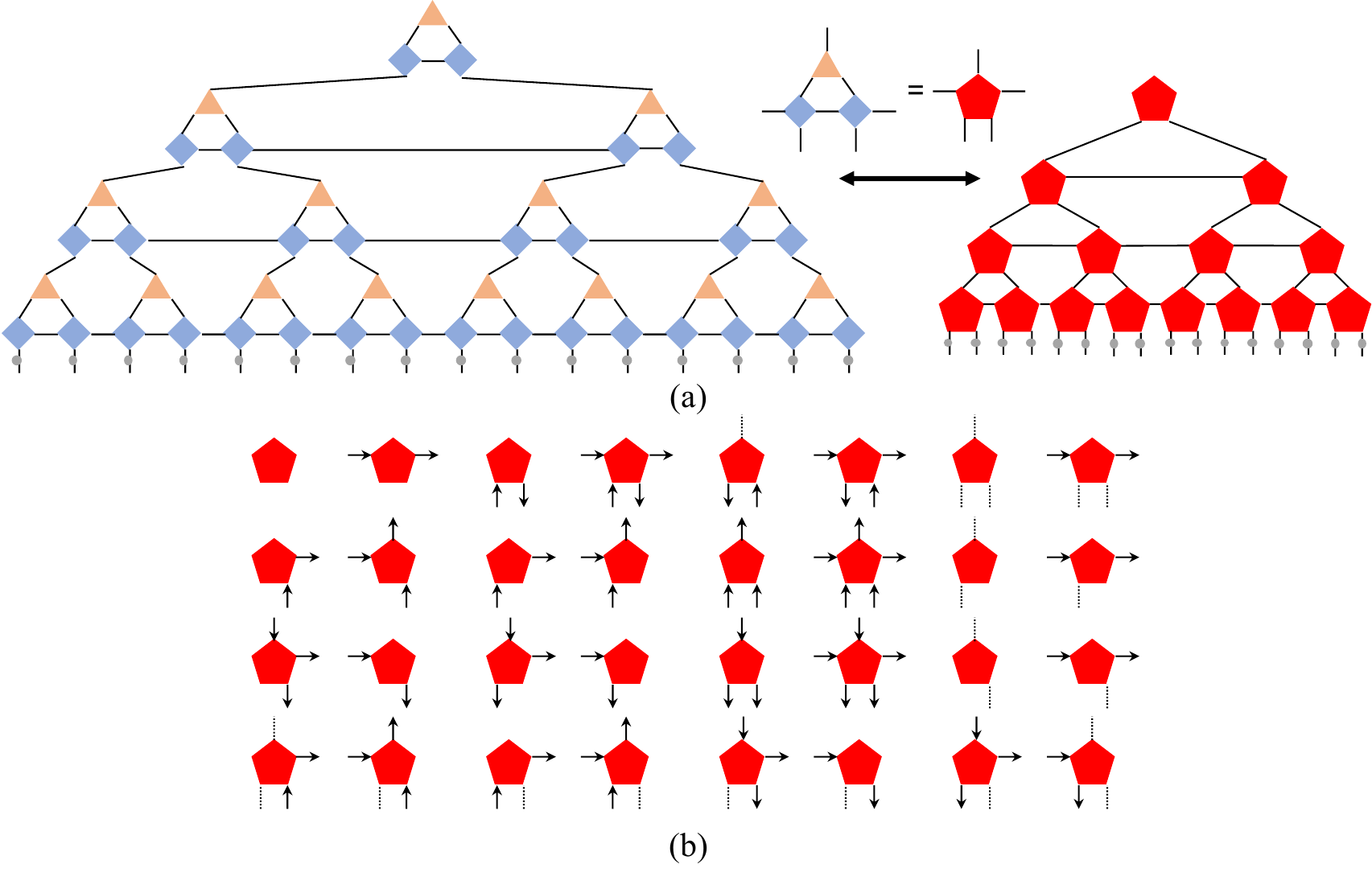}
    \caption{(a) Contracting three tensors of the original Motzkin TN gives a uniform pentagonal TN with a more intuitive definition of the elementary tensor. (b) The 32 non-vanishing entries of the rank-5 tensor in the bulk of the TN. The tensors on the boundary of the network has rank two or four, depending on its location, and is only non-vanishing for the corresponding compatible configurations. The small gray circles on the external legs of the network project out the index configuration corresponding to the dotted lines (and empty lines for the Fredkin case).}
    \label{fig:penTN}
\end{figure*}

The GSs of Motzkin and Fredkin chains are superpositions of classical random walk configurations, which makes it convenient to express them by contraction of local tensors. A hierarchical TN representation for both of them was constructed in Ref.~\cite{Alexander2021exactholographic}, with building blocks of rank-3 and -4 tensors as shown on the left panel of \cref{fig:penTN} (a). Here, we contract one rank-3 with two rank-4 tensors below it, according to the list in Fig.~11 and 12 of the Appendix D and E of Ref.~\cite{Alexander2021exactholographic} to form an elementary tensor of rank-5. The resulting network has only one building block seen on the right panel of \cref{fig:penTN} (a) for a system of size $L=8$. The specification of the nonvanishing entries of the tensors in Ref.~\cite{Alexander2021exactholographic} translates to the definition in \cref{fig:penTN} (b), where all listed tensor components have the same value 1 and all the rest are of value 0. Since each horizontal leg of the rank-5 tensor has bond dimension two, while each of its vertical leg is of dimension 4, there are 32 out of the $2^2\cdot 4^3=256$ possible combinations that give non-vanishing elements of the tensor, obviously a very sparse one. Each of the external legs on the bottom layer is attached to a projection operator represented by the gray dots that projects out the unphysical degrees of freedom denoted by the dotted line (and the empty leg for the spin-$\frac{1}{2}$ Fredkin model described by the same TN). 

The three legs on the left and bottom of a pentagon can be labeled as the input indices, and the top and right ones the output. Ignoring the dotted lines for the moment, an arrow flowing inward (resp.~outward) is considered $+1$ (resp.~$-1$) for the input legs, and the opposite for the output, while an empty leg is treated as 0. Then each pentagonal tensor corresponds to a binary addition, with the top leg encoding the carry digit. This represents a coarse-graining mapping from the spins in the level below to the level above. Two up- or down-spins are of course mapped to up and down respectively, and one up one down is mapped to spin zero. But what one up one zero or one down one zero maps to is dependent on the leftover spin from the string of tensors on the same level to the left. The role of the dotted line is to keep track of reverse ordered up-down spin neighboring pairs. This is important for the GS to be a superposition of only spin configurations that correspond to Motzkin paths, namely to always have at least as many up spins as down spins to the left at any point along the chain. Therefore, two dotted lines merge into one, and a dotted line disappears in the level above if the horizontal legs of the tensor has nonzero values. 

\begin{figure}[ht]
    \centering
    \includegraphics[width=0.9\linewidth]{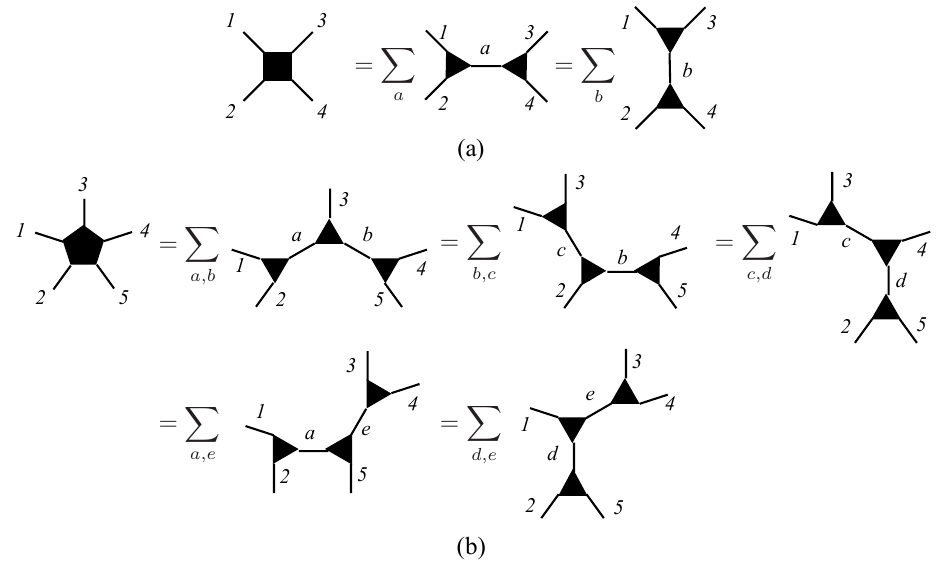}
    \caption{(a) Two different ways to decompose a rank-4 tensors. (b) Five different decompositions of a 5-leg tensor.}
    \label{fig:penid}
\end{figure}
\begin{figure}[ht]
    \centering
    \includegraphics[width=0.9\linewidth]{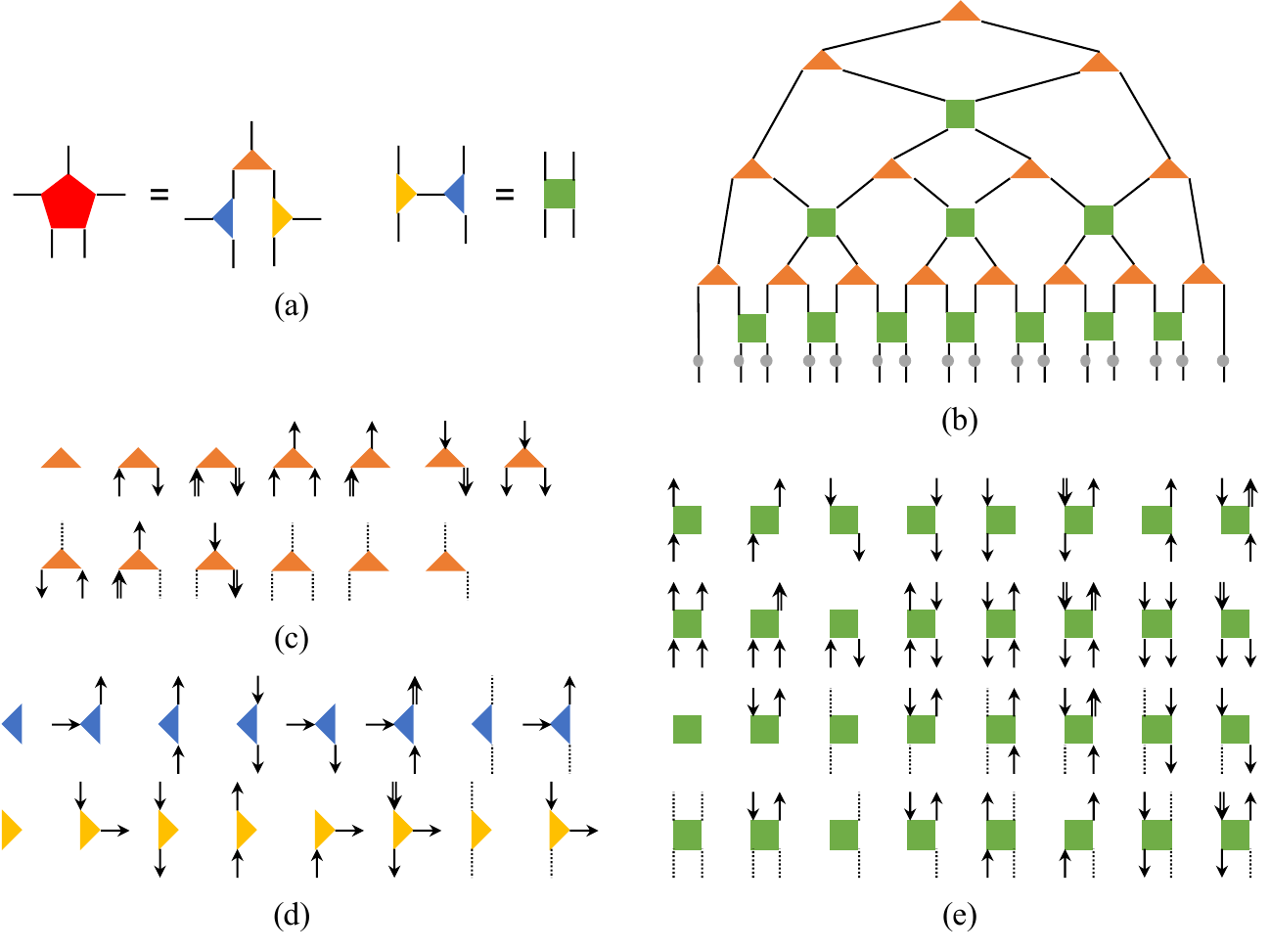}
    \caption{(a) Decomposition of a rank-5 tensor into the product of three rank-3 tensors, and recombination of two of the rank-3 tensors into a rank-4 tensor. (b) Equivalent representation of the pentagonal TN in \cref{fig:penTN} (a) using rank-3 and 4 tensors. (c)-(e) List of tensor configurations for the four types of tensors with value 1. Tensor entries for the rest of the configurations all have value 0.}
    \label{fig:tensors}
\end{figure}

In order to put the TN of \cref{fig:penTN} (a) into the geometry of a MERA, the rank-5 tensor must first be decomposed into the contraction of three rank-3 tensors. In general, there are five ways to do that without braiding the external legs, as illustrated in \cref{fig:penid}. For symmetry considerations, we will work with the one shown on the left panel of \cref{fig:tensors} (a). The three resulting rank-3 tensors are given by those in \cref{fig:tensors} (c) and (d) having value 1, and 0 otherwise. Two of them (\cref{fig:tensors} (d)) can be combined between neighboring tensors following the right panel of \cref{fig:tensors} (a) to form the rank-4 tensor in \cref{fig:tensors} (e). The difference between the TN in \cref{fig:tensors} (b) and the MERA is that the 4-leg tensor is not unitary (although the 3-leg tensor in \cref{fig:tensors} (c) can be made isometric by letting the upper legs of the tensors in \cref{fig:tensors} (d) absorb a constant factor.) Unlike MERA, in addition to not being unitary or isometric, the tensors here also implement coarse-graining with both the rank-3 and -4 tensors. 

The translational symmetry along the physical external legs of a hierarchical TN like MERA is broken even for infinitely large systems without boundaries. At the first level of the TN, the discrete translational invariance is broken into translations by two lattice spacings, which is further broken at the next level to translations by four lattice spacings, and so on so forth. So the hierarchical structure of the TN completely breaks the translational invariance. This is perhaps one of the reasons that so far there has not been an exactly solvable Hamiltonian that has a GS described exactly by the MERA. 

\section{Derivation of \texorpdfstring{$\mathcal{T}^{k_{1}k_{2}}_{S^{z}}$}{T} at \texorpdfstring{$q=1$}{q=1}}\label{sec:exp_val_derivation}
In this appendix we derive the closed form expression of $\mathcal{T}^{k_{1}k_{2}}_{S^{z}}$ in the $q = 1$ case needed to arrive at the final expression for $\langle S^{z}_{r}\rangle$ in \cref{eq: Sz_expectation_expression}. We start with $\mathcal{T}^{k_{1}k_{2}}_{S^{z}}$, which is obtained by inserting the matrix representation for $S^{z}$ for $O$ in \cref{eq: To_k1k2_general}. Using that the eigenvectors $|k_{i}\rangle$ of the maximum dimension matrix block are given by \cref{eq: eigenToeplitz} gives
\begin{equation}
    \begin{split}
        \mathcal{T}^{k_{1}k_{2}}_{S^{z}} &= \langle k_{1}|\mathcal{T}_{S^{z}}|k_{2}\rangle,\\
        &=\frac{2S}{d_{\text{max}}}\sum^{d_{\text{max}}}_{j, l = 1}\sin{\left(\frac{\pi k_{1}j}{d_{\text{max}} +1}\right)}\left(\delta_{j, l-1}-\delta_{j, l+1}\right)\sin{\left(\frac{\pi k_{2}l}{d_{\text{max}} +1}\right)}.\\
    \end{split}
    \label{eq: TSz_k1k2}
\end{equation}
Note that $(\mathcal{T}_{S^{z}})_{jl} =S(\delta_{j, l-1}-\delta_{j, l+1})$, with $S = \frac{1}{2}$ for the Fredkin chain and $S = 1$ for the Motzkin chain. This is seen by inserting the matrix representation for $S^{z}$ for $O$ in \cref{eq: To_k1k2_general}. Since we are interested in the large $L$ limit, the following simplification $d_{\text{max}} +1 \approx d_\text{max} \approx L$ is made. \Cref{eq: TSz_k1k2} then becomes

\begin{equation}
\begin{split}
    \mathcal{T}^{k_{1}k_{2}}_{S^{z}} &=\frac{4S}{L}\sin{\left(\frac{\pi k_{2}}{L}\right)}\sum^{L}_{j=1}\sin{\left(\frac{\pi k_{1}j}{L}\right)}\cos{\left(\frac{\pi k_{2}j}{L}\right)},\\
    &=\frac{2S}{L}\sin{\left(\frac{\pi k_{2}}{L}\right)}\sum^{L}_{j=1}\left[\sin{\left(\frac{\pi(k_{1} +  k_{2})j}{L}\right)}+\sin{\left(\frac{\pi(k_{1} -  k_{2})j}{L}\right)}\right],
\end{split}
    \label{eq: OperatorOverlapSz}
\end{equation}
where we have used product-to-sum trigonometric identities to obtain the final expression. The sums in \cref{eq: OperatorOverlapSz} are performed using Lagrange's trigonometric identities, to obtain the closed form expression. Using that $k_{1}$ and $k_{2}$ are integers, we obtain \cref{eq: OperatorOverlapSz_final}. Note that for even $k_{1} + k_{2}$, the expression vanishes. Inserting this expression into \cref{eq: ExpectationValueGeneralFinal} gives \cref{eq: Sz_expect_start}.\\

\section{Evaluation of the integrals}
\label{sec:IntegralEval}

In this appendix we evaluate the double integral $I$ in \cref{eq: Sz_expect_final_integral} in the limit $1\ll r\ll L$. For the Fredkin case, the main contribution of the integral comes from the region close to $x = 0, \pi$ and $y = 0, \pi$. We therefore split the double integral in four terms, each capturing one of the maxima. This gives
\begin{equation}
    I(r)= \left[\int^{\frac{\pi}{2}}_{0}\int^{\frac{\pi}{2}}_{0}dxdy\dots+ \int^{\frac{\pi}{2}}_{0}\int^{\pi}_{\frac{\pi}{2}}dxdy\dots + \int^{\pi}_{\frac{\pi}{2}}\int^{\frac{\pi}{2}}_{0}dxdy\dots+ \int^{\pi}_{\frac{\pi}{2}}\int^{\pi}_{\frac{\pi}{2}}dxdy\dots\right],
    \label{eq: integral_split}
\end{equation}
where the dots refer to the integrand of \cref{eq: Sz_expect_final_integral}. For each of the four terms, we perform a saddle-point approximation, by expanding the integrand around the point of maximum value. For the first term in \cref{eq: integral_split}, $I_{1}$, the integrand is sharply peaked around $x = 0$ and $y = 0$. The saddle-point approximation is then performed by Taylor expanding the integrand in the vicinity of $x = 0$ and $y = 0$, which gives
\begin{equation}
    I_{1}(r) = 4\iint^{\pi/2}_{0}dxdy\frac{x^{2}y^{2}}{x^{2}- y^{2}}e^{-\frac{(r-1)x^{2}}{\lambda_{*}}}e^{-\frac{(2L-r)y^{2}}{\lambda_{*}}}. 
    \label{eq: Sz_I_1_integral}
\end{equation}
Here we have used that $\cot{(\frac{x\pm y}{2})}\approx \frac{2}{x\pm y}$ for $x$ and $y$ close to 0 and Taylor expanded the $\cos$-functions followed by an exponential approximation. To proceed, notice that the exponential functions quickly vanish as $x$ and $y$ increase. The characteristic width of the exponential functions are $\sim \lambda_{*}/\sqrt{r-1}$ and $\sim \lambda_{*}/\sqrt{2L-r}$, both fully contained within the integration limits when $1\ll r\ll L$. We therefore extend the limit to infinity without affecting the result significantly
\begin{equation}
    I_{1}(r)
    \approx 4\iint^{\infty}_{0}dxdy \frac{x^{2}y^{2}}{x^{2}- y^{2}}e^{-\frac{(r-1)x^{2}}{\lambda_{*}}}e^{-\frac{(2L-r)y^{2}}{\lambda_{*}}}.
\end{equation}
Next, we perform substitutions $u = x\sqrt{\frac{2(r-1)}{\lambda_{*}}}$ and $v = y\sqrt{\frac{2(2L-r)}{\lambda_{*}}}$, giving
\begin{equation}
\begin{split}
    I_{1}(r)&=\frac{4\lambda_{*}^{2}}{2^{2}\sqrt{(r-1)(2L-r)}}\iint^{\infty}_{0}dudv \frac{u^{2}v^{2}}{(2L-r)u^{2}- (r-1)v^{2}}e^{-\frac{u^{2}}{2}}e^{-\frac{v^{2}}{2}},\\
    &= \frac{\lambda_{*}^{2}}{(2L)^{3/2}\sqrt{(r-1)(1-\frac{r}{2L})}}\iint^{\infty}_{0}dudv \frac{u^{2}v^{2}}{(1-\frac{r}{2L})u^{2}- \frac{r-1}{2L}v^{2}}e^{-\frac{u^{2}}{2}}e^{-\frac{v^{2}}{2}}.
\end{split}
\end{equation}
In the limit $r\ll L$, it becomes an elementary integral
\begin{equation}
\begin{split}
    I_{1}(r)\approx \frac{\lambda_{*}^{2}}{(2L)^{3/2}\sqrt{r-1}}\iint^{\infty}_{0}dudv v^{2}e^{-\frac{u^{2}}{2}}e^{-\frac{v^{2}}{2}}= \frac{\pi \lambda_{*}^{2}}{2(2L)^{3/2}\sqrt{r-1}}.
\end{split}
\label{eq: I_1}
\end{equation}

Next up is the second term in \cref{eq: integral_split}, $I_{2}$, where we perform the expansions around $x = 0$ and $y = \pi$. For the Motzkin chain, this produces a term that vanishes as $3^{r-2L}$. For the Fredkin chain this gives 
\begin{equation}
\begin{split}
        I_{2}(r) &= (-1)^{2L-r} \int^{\pi/2}_{0}\int^{\pi}_{\pi/2}dxdy (-x^{2})(\pi - y)^{2}e^{-\frac{(r-1)x^{2}}{2}}e^{-\frac{(2L-r)(\pi-y)^{2}}{2}},\\
        &=(-1)^{r+1} \int^{\pi/2}_{0}\int^{\pi/2}_{0}dxdy x^{2}y^{2}e^{-\frac{(r-1)x^{2}}{2}}e^{-\frac{(2L-r)y^{2}}{2}},
\end{split}
\label{eq: I_2_integral}
\end{equation} following the same steps as for $I_{1}$. Note that the $(-1)^{p}$ term results from expanding $\cos^{p}(y)$ around $y = \pi$. To obtain the second line in \cref{eq: I_2_integral}, we have performed the substitution $y^{\prime} = \pi-y$ and switched upper and lower limits. Following the same steps as for $I_{1}$, \cref{eq: I_2_integral} is evaluated as 
\begin{equation}
\begin{split}
    I_{2}(r) 
    &\approx (-1)^{r+1}\iint^{\infty}_{0}dxdy x^{2}y^{2}e^{-\frac{(r-1)x^{2}}{2}}e^{-\frac{(2L-r)y^{2}}{2}},\\
    &= \frac{(-1)^{r+1}}{(2L)^{3/2}(r-1)^{3/2}}\iint^{\infty}_{0}dudv u^{2}v^{2}e^{-\frac{u^{2}}{2}}e^{-\frac{v^{2}}{2}},\\
    &=\frac{(-1)^{r+1}\pi}{2(2L)^{3/2}(r-1)^{3/2}}.\\
\end{split}
\label{eq: I_2}
\end{equation}

For the Fredkin chain, the last two terms in \cref{eq: integral_split}, $I_{3}$ and $I_{4}$, are found to be identical to $I_{2}$ and $I_{1}$ respectively. For $I_{3}$ this is seen by expanding around $ x = \pi$ and $y = 0$ and then substituting $x^{\prime} = \pi - x$. For $I_{4}$ this is seen by expanding around $ x = \pi$ and $y = \pi$ and performing the same substitution for both $x$ and $y$. Thus, we have found
\begin{equation}
   I_{\text{Fredkin}}(r)  \approx2\left[I_{1}(r) + I_{2}(r)\right].
    \label{eq: final_integral_result_F}
\end{equation}

For the Motzkin case, because the spectrum of the transfer matrix is asymmetric about 0, its largest eigenvalues are located in the region where both $x$ and $y$ are near 0. So there is no need to split the integral, and it is safe to extend the upper boundary of the integrals $\pi$ to infinity. So the result is identical with \cref{eq: I_1} for the Fredkin case, differing only in the value of $\lambda_*$.

\section{Domain wall thickness from mean-field theory}\label{sec:MFT}

In this appendix, we derive the relation between domain wall thickness and correlation length from the Ginzburg-Landau (G-L) free energy
\begin{equation}
    f(\phi(l))=(\partial_l\phi)^2+b\phi^2+c\phi^4,
\end{equation}for the field $\phi(l)=\langle S^z_l\rangle$, as its spatial average is vanishing due to parity symmetry. The G-L equation
\begin{equation}
    \partial_l^2\phi=b\phi+2c\phi^3
\end{equation} has no explicit dependence on the variable $l$, so it has the first integral
\begin{equation}
    I=f-\partial_l\phi \frac{\partial f}{\partial (\partial_l\phi)}=-(\partial_l\phi)^2+b\phi^2+c\phi^4,
\end{equation}satisfying $\partial_l I=0$. Near the two boundaries, we have the two stationary solutions 
\begin{equation}
    \phi(L\mp L)=\pm\sqrt{\frac{-b}{2c}}\equiv \pm 1, 
\end{equation} and $\partial_l\phi|_{l=0,2L}=0$ for $b<0$. 
Using the conservation of the first integral, we have
\begin{equation}
    \partial_l \phi= \sqrt{c}(\phi^2-1), 
\end{equation}which integrates to 
\begin{equation}
    \phi(l)=\tanh\left(\sqrt{c}(L-l)\right).
\end{equation}On the other hand, the correlation length is given by the inverse mass $\xi=\sqrt{\frac{1}{-2b}}$ for the ordered phase $b<0$. So we see the domain wall thickness predicted by the G-L theory is exactly twice the correlation length.

\end{appendix}

\bibliography{penta.bib}

\end{document}